\documentclass[aps,pra,reprint,showpacs,twocolumn,amsmath,amssymb]{revtex4}
\usepackage{epsfig}
\usepackage{ifpdf}
\usepackage{graphicx}
\usepackage{amssymb,amsmath,lineno,mathbbol,upgreek}
\usepackage[tight]{subfigure}
\usepackage{mathrsfs}
\usepackage{bbm}
\usepackage{color}
\usepackage{slashed}
\usepackage{natbib}

\usepackage{calligra}
\DeclareMathAlphabet{\mathcalligra}{T1}{calligra}{m}{n}
\DeclareFontShape{T1}{calligra}{m}{n}{<->s*[2.2]callig15}{}

\newcommand{\hn}{{\boldsymbol{\hat{\bf{n}}}}}
\newcommand{\hx}{{\boldsymbol{\hat{\bf{x}}}}}
\newcommand{\hy}{{\boldsymbol{\hat{\bf{y}}}}}

\usepackage{float}
\usepackage{afterpage}
\usepackage{float}
\usepackage{lipsum}

\usepackage{calligra}
\DeclareMathAlphabet{\mathcalligra}{T1}{calligra}{m}{n}
\DeclareFontShape{T1}{calligra}{m}{n}{<->s*[2.2]callig15}{}

\pdfoutput=1

\pacs{03.75.Lm, 67.85.-d, 05.45.-a, 03.65.Pm}

\begin{document}

\title{The nonlinear Dirac equation: Preparation and stability of relativistic vortices in Bose-Einstein condensates}

\author{L. H. Haddad$^{1}$, K. M. O'Hara$^2$, and Lincoln D. Carr$^{1,3}$}
\affiliation{$^1$Department of Physics, Colorado School of Mines, Golden, CO 80401,USA    \\$^2$Department of Physics, Pennsylvania State University, University Park, Pennsylvania 16802-6300, USA\\$^3$Physikalisches Institut, Universit\"at Heidelberg, D-69120 Heidelberg, Germany}
\date{\today}

\begin{abstract}
 We propose a detailed experimental procedure for preparing relativistic vortices, governed by the nonlinear Dirac equation, in a two-dimensional Bose-Einstein condensate (BEC) in a honeycomb optical lattice. Our setup contains Dirac points, in direct analogy to graphene. We determine a range of practical values for all relevant physical parameters needed to realize relativistic vortices in a BEC of $^{87}\mathrm{Rb}$ atoms. Seven distinct vortex types, including Anderson-Toulouse and Mermin-Ho skyrmion textures and half-quantum vortices, are obtained, and their discrete spectra and stability properties  are calculated in a weak harmonic trap. We predict that most vortices are stable with a lifetime between $1$ and $10$ seconds. 
\end{abstract}

\maketitle

\section{Introduction}

Multi-component Bose-Einstein condensates (BECs) present an ideal setting for studying complex vortex structures~\cite{Kawaguchi2011}. Such vortices allow for topologically intriguing configurations ranging from skyrmions to knots~\cite{Ketterle2003,Ueda2008,Zhou2008}. The usual method for adding a spinor structure to a BEC relies on hyperfine degrees of freedom or different atomic species. Instead, we use the band structure and linear dispersion relation around the Dirac points at the Brillouin zone edge of a honeycomb optical lattice to realize a four-component Dirac spinor, in direct analogy to graphene~\cite{Geim:2007}. This gives us both pseudospin as well as a relativistic structure. To accomplish this, we propose starting with a BEC of weakly interacting alkali metal atoms in the lowest Bloch state of a quasi-two-dimensional (quasi-2D) honeycomb optical lattice, then using Bragg scattering to populate Bloch states at the two inequivalent Dirac points, followed by the application of a Laguerre-Gaussian laser beam to deliver a net angular momentum to the BEC which excites a plethora of vortex structures. The vortices we obtain are solutions of the nonlinear Dirac equation (NLDE), whose stability is determined by the relativistic linear stability equations (RLSE)~\cite{haddad2009,haddad2011}. Our work on the NLDE+RLSE system opens up the field of relativistic simulations in BECs at velocities ten orders of magnitude slower than the speed of light.

  In this article we combine the study of Dirac points with superfluid vortices, an environment reminiscent of particle physics models where relativistic vortices are commonplace~\cite{Nielsen1973,Seiberg1994}. Stability of a BEC at the Dirac points presents a challenge, since Bloch states there have finite crystal momentum and nonzero energy. We handle this problem by introducing an intermediate asymmetry between the A and B sublattice potential depths which opens up a mass gap. Using a gap enables us to construct initial and final Bloch states, $\psi_{A, \mathbf{0}}$ and $\psi_{A, \mathbf{K}}$ (with Dirac point momentum $\mathbf{K}$), as superpositions of the two degenerate states at the Dirac point with velocities $c_l$ and $- c_l$, respectively. This produces a state with group velocity equal to zero, relative to the lattice. Stationarity of the BEC with respect to the lattice and the lab frame is significant experimentally, since the BEC can remain confined in an external trapping potential indefinitely and does not suffer from dynamical instabilities associated with relative motion between the BEC and lattice. The end result is a metastable state in which thermal losses can be managed by maintaining the system at very low temperatures. For realistic experimental parameters our relativistic vortices are stable for up to 10 seconds, as long or longer than the lifetime of typical BECs.

Our physical setting begins with a BEC tightly confined in one direction and loosely confined in the other two directions. More precisely stated, magnetic trapping along the $z$-direction is such that excitations along this direction have much higher energy, by at least an order of magnitude, compared to the lowest excitations in the $x$ and $y$-directions. Thus, an important step is to calculate the precise renormalization of all relevant physical parameters when transitioning from the standard 3D BEC to a quasi-2D system. In addition to this step we also account for a renormalization due to the presence of the optical lattice potential which introduces an additional length scale from the lattice constant. We point out that microscopically the BEC obeys the three-dimensional nonlinear Schr\"odinger equation and we consider temperatures well below the BKT transition energy associated with two-dimensional systems. Nevertheless, throughout our work we often use ``2D'' for brevity, keeping in mind the quasi-2D picture. Condensation at Dirac points of the honeycomb lattice requires additional techniques beyond ordinary condensation, which we detail in this article. In addition to the fields needed to construct the lattice one requires both walking and stationary standing wave optical potentials to respectively Bragg scatter atoms form the ground state (zero crystal momentum) to the Dirac point and also between inequivalent Bragg points.

This article is organized as follows. In Sec.~\ref{sec:PhysicalParameters}, we discuss physical parameters, constraints, and regimes. In Sec.~\ref{sec:LatticeConstruction}, we present methods for constructing the honeycomb optical lattice. In Sec.~\ref{sec:DiracPointPrep}, we propose steps for preparing a BEC at a Dirac point that has amplitude on only one of the sublattices. Section~\ref{sec:SublatticeTransfer} describes a process to coherently transfer atoms between sublattices. In Sec.~\ref{sec:BraggScatter}, we explain a procedure for coherently transferring the BEC between inequivalent Dirac points, the final step needed to control the amplitude of all four components of the Dirac spinor. Section~\ref{sec:VortexSolutions} presents vortex solutions of the NLDE including stability analysis. In Sec.~\ref{sec:CreateVortex}, we explain how to excite NLDE vortices by modifying the procedure described in Sec.~\ref{sec:SublatticeTransfer} to include co-propagating Gaussian and Laguerre-Gaussian laser beams which transfer angular momentum to the BEC.  In Sec.~\ref{sec:Conclusion}, we conclude.

\section{Physical parameters and constraints}
\label{sec:PhysicalParameters}

\begin{table*}[]
\resizebox{18cm}{!}{
\begin{tabular}{lllll}
\multicolumn{5}{c}{} \\
 \hline \hline
Parameter &    Symbol/Definition & Constraint & \hspace{2pc} Value & \hspace{2pc} Range  \hspace{1pc}   \\
\hline
(a)  Temperature  &\; $T$ & $  \ll \hbar \omega_z $    & \; $2.5 \, \mathrm{nK}$ &  \; $ \sim 100\, \mathrm{pK} < T \lesssim 80 \, \mathrm{nK}$  \\
(b)  Chemical potential  &\; $\mu $ & $  \ll \hbar \omega_z $      &\;  $2.36 \, \mathrm{nK}$ &\;  $ <  4.10 \, \mathrm{nK} $  \\
(c)  Transverse oscillator length  & \;  $L_z =( \hbar /M \omega_z)^{1/2} $ \; & $  \ll R_\perp$ & \;  $1.50\, \mathrm{\mu m}$ \;  & \; $ <   3.0 \, \mathrm{\mu m}$  \\
(d)  Healing length  & \; $\xi = 1/\sqrt{8 \pi \bar{n} a_s} $ \; & $ \lesssim L_z$  & \;  $1.10 \, \mathrm{\mu m}$ \; & \; $\lesssim  1.50  \, \mathrm{\mu m}$   \\
(e) Effective speed of light  & \;  $c_l = t_h a \sqrt{3}/2 \hbar$ \; & $ <  c_{s,\mathrm{2D}}$ &\;   $5.31 \times 10^{-2}\, \mathrm{cm}/\mathrm{s} $ \; & \; $< 5.40 \times 10^{-2} \, \mathrm{cm}/\mathrm{s}$  \\
(f)  Dirac nonlinearity  & \;  $U =     L_z    \, g \, \bar{n}^2   \,3   \sqrt{3}\,  a^2 /8$ \;   & $ \ll t_h, \, \mu$  &\;   $1.07 \, \mathrm{nK}$ \; & \;  $ <  2.36  \, \mathrm{nK}$  \\
(g)  Quasi-particle momentum  & \;  $ k =  p / \hbar $ \; &  $  \ll  \sqrt{8}/a$    &  \; $6.27 \times 10^{2} \,  \mathrm{cm}^{-1}$ \;    &  \; $6.27 \times 10^{2} \,  \mathrm{cm}^{-1} \lesssim  k \ll 5.66 \times 10^{4} \, \mathrm{cm}^{-1}$   \\
(h)  Dirac healing length  & \;  $\xi_\mathrm{Dirac} = t_h a \sqrt{3}/2 U$ \; &  $\gg a , \; \ll R_\perp$ &\;   $3.80 \, \mathrm{\mu m}$ \; &\;  $0.50  \, \mathrm{\mu m}  \ll \xi_\mathrm{Dirac}    \ll  50.0\, \mathrm{ \mu m}$     \\
(i) Lattice depth  & \;  $V_0$ \; &  $\gg E_R$ &\;   $10.1 \, \mathrm{\mu K}$ \; &\;  $0.79 \, \mathrm{\mu K} < V_0 < 10.1 \, \mathrm{\mu K} $     \\
\hline \hline
\end{tabular}    }
\caption{\emph{Physical parameters and constraints for the NLDE typical for a BEC of $^{87}$Rb atoms.} (a,b) Relative energies for the 3D to quasi-2D dimensional reduction, with the vertical trap oscillator energy $\hbar \omega_z$. (c,d,h) Relative lengths for the 3D to quasi-2D dimensional reduction. (e) Landau criterion imposed to avoid dynamical instabilities, where the quasi-2D speed of sound in the continuum $c_{s,\mathrm{2D}} \equiv \sqrt{3 g \bar{n}/2M}= 5.90 \times 10^{-2}\, \mathrm{cm}/\mathrm{s}$. Note that the factor of $\sqrt{3/2}$ comes from integrating over the vertical dimension. (f) The weakly interacting and superfluid (not Mott insulating) regime. (g) The linear Dirac cone approximation which requires that quasi-particle momenta $\hbar k$ remain small compared to the Dirac point momentum $\hbar K$. (h) Long-wavelength limit, which sets the scale for the quasi-2D Dirac healing length. (i) The lowest-band and tight-binding approximation. For the values in the table, we use the ratio of lattice depth to recoil energy $V_0/E_R = 16$, lattice constant $a= 2 \lambda_L/3 = 0.28 \, \mathrm{\mu m}$, and planar trap radius $R_\perp = 100 \, a$, average particle density $\bar{n} = 5.86 \times  10^{18}\, \mathrm{m}^{-3}$, hopping energy $t_h = 16.8 \, \mathrm{nK}$, and atomic mass of $^{87}$Rb.} \label{table1}  
\end{table*}

Relativistic vortices are realized in the emergent nonlinear Dirac background, in the long wavelength limit of a quasi-2D honeycomb lattice obtained by tightly constraining the system  in one spatial dimension (the z-direction). Thus, microscopically, the BEC obeys the three-dimensional nonlinear Schr\"odinger equation, but vibrational excitations in the z-direction are avoided. The usual 3D BEC parameters are renormalized, once for the dimensional reduction~\cite{Carr2000}, and again after integrating over the lattice Wannier functions and going to long wavelengths. Consequently, NLDE physics is only experimentally realizable in practice when several energy and length constraints are satisfied. We list these constraints in Table~\ref{table1} along with their mathematical definitions. For our calculations, we use the semiclassical estimate~\cite{Lee:2009} of the hopping parameter $t_h  \equiv 1.861 \left(V_0/E_R\right)^{3/4} E_R\, \mathrm{exp}\! \left( -1.582 \sqrt{V_0/E_R} \right)$. It is helpful to consolidate the constraint inequalities to arrive at expressions relating the temperature $T$ and length scales of the system, $a_s$, $a$, $d$, $L_z$, and $R_\perp$:
\begin{eqnarray}
\hspace{-2pc}&&1 \lesssim \left( \frac{8 \pi   a_s}{d^3}  \right)^{3/2}\! \!  L_z^3 <  \frac{2^5 \sqrt{2}\, \pi^{1/2} (d^3a_s)^{1/2}}{3 \sqrt{3} \, a^2 \left[ 1 +  \pi a/(4 \sqrt{2} R_\perp) \right] } \, , \label{reducedbounds1} \\
\hspace{-2pc}&&T  <  \hbar^2/k_B M L_z^2  \, , \label{reducedbounds2}
\end{eqnarray}
 where $d$ is the average inter-particle distance defined in terms of the particle density $d= \bar{n}^{-1/3}$. All other quantities are defined in Table~\ref{table1}. The temperature $T$ in Eq.~(\ref{reducedbounds2}) depends indirectly on the ratio $V_0/E_R$ through $\bar{n}$. We can get an idea of how the particle density affects $T$ by evaluating the inequalities for different values of $\bar{n}$ while fixing $V_0/E_R =16$. For example, $\bar{n} = 10^{16}\, \mathrm{m}^{-3}$ gives $26.259 \, \mathrm{\mu m} \lesssim L_z < 86.934 \, \mathrm{\mu m}$ and $T < 8.17 \times 10^{-3} \, \mathrm{nK}$, whereas for $\bar{n} = 10^{20} \, \mathrm{m}^{-3}$ we find $  0.187  \, \mathrm{\mu m} \lesssim L_z <  0.263  \, \mathrm{\mu m}$ and $T <  162  \, \mathrm{nK}$. From this we see that a practical value for $T$ requires that densities be considerably larger than $10^{16} \, \mathrm{m}^{-3}$, a consequence of the additional constraints in Eqs.~(\ref{reducedbounds1})-(\ref{reducedbounds2}). We next address the required constraints in detail and explore the conditions under which each is satisfied.

 In order to obtain an effectively 2D system, the vertical oscillator length must be much smaller than the trap size along the direction of the plane of the condensate. Hence, for an effectively 2D system the required length constraint implies the condition $L_z \ll R_\perp$. Taking $R_\perp \approx 100\,  a$ (a typical condensate size), and using a realistic value for the vertical oscillator length (Table~\ref{table1}), we obtain $L_z = 5.36 \, a$, which satisfies the constraint. Moreover, we require a healing length close to or less than the transverse oscillator length. With $\xi = 1.10 \, \mathrm{\mu m}$ and $L_z = 1.50 \,  \mathrm{\mu m}$, we find that this condition holds. Another necessary condition for realizing the NLDE in the laboratory is that the healing length (defined in the effective Dirac theory) must be much larger than the lattice constant. The long-wavelength limit is thus defined by $\xi_\mathrm{Dirac} /a \gg 1$, for which we find that $\xi_\mathrm{Dirac} /a = 13.57$.

The Landau criterion for the effective velocities in the BEC is required in order to avoid the instabilities associated with propagation faster than the sound speed in the condensate. This condition demands that the effective speed of light is less than the 2D renormalized speed of sound. Stated mathematically, the Landau criterion requires that $c_l / c_{s, \mathrm{2D}}  <1$. Using the definitions for the effective speed of light and the sound speed consistent with Table~\ref{table1}, we compute $c_l / c_{s, \mathrm{2D}}  =  0.90$, which satisfies the inequality.

The NLDE and RLSE are derived for a weakly interacting Bose gas. This ensures both the stability of the condensate as well as the effective nonlinear Dirac mean field description. We then require the interaction energy to be significantly less than the total energy of the system. The energy constraints may be stated as $\mu, k_B T \ll \hbar \omega_z$. We can solve the NLDE for the lowest excitation to obtain an expression for the chemical potential $\mu = \hbar c_l k + U |\Psi|^2$~\cite{haddad2011}. Next, we evaluate this expression using the lowest excitation in a planar condensate of radius $R \approx 100 a$, which has wavenumber $k \approx \pi/2R= 2.86 \times 10^4 \, \mathrm{m}^{-1}$. The interaction $U$ is computed using the quasi-2D renormalized interaction in Table~\ref{table1} for the binary interaction $g$  and mass $M$ pertaining to a condensate of $^{87}\mathrm{Rb}$ atoms. Finally, for a uniform condensate we take $|\Psi|^2 = 1 $ and the constraint on the chemical potential becomes $\mu = 2.36 \, \mathrm{nK} < 22.17\, \mathrm{nK}$, which is satisfied. For the temperature, we require $ T\ll \hbar \omega_z /k_B$. Using the data in Table~\ref{table1} for the vertical oscillator frequency, we obtain the upper bound for the temperature $T \ll  22.17 \,  \mathrm{nK}$. This is a reasonable requirement given that BEC occurs for $T$ in tens or hundreds of nanoKelvins or as low as picoKelvins. 

For a condensate in the regime where the NLDE description is valid, we require that the linear approximation to the exact dispersion remain valid. As in the case of graphene, large deviations from the Dirac point induce second order curvature corrections to the dispersion. Thus, we must quantify the parameter restrictions which allow for a quasi-relativistic interpretation. To quantify this, we expand the exact dispersion near the Dirac point to obtain $\mu(k) =  U \pm t_h \left(  a \sqrt{3}k /2+ a^2 k^2/8 - a^3 \sqrt{3} k^3/48 + ...   \right)$,  where $k$ is the small momentum parameter which measures the deviation away from the Dirac point. Notice that the first order term gives the linear dispersion of the Dirac equation while higher order corrections describe the bending of the band structure as we move away from the Dirac point. The second order term tells us that the NLDE description is valid as long as $a k /\sqrt{8} \ll 1$, which determines a lower bound on the wavelength for fluctuations of the condensate away from the Dirac point: $\lambda_\mathrm{min} \gg (2 \pi/\sqrt{8}) a$. The requirement of maintaining the linear dispersion then places an additional constraint on the chemical potential, namely that $ |\mu| \ll  U + 6 t_h \simeq 101.9 \, \mathrm{nK}$. Using the value for the chemical potential found earlier, we see that $\mu = 2.36 \,  \mathrm{nK}  \ll  101.9 \, \mathrm{nK}$. Finally, since we are treating the case of weak short range interactions at very low temperatures, the lowest band approximation is sufficient to describe the physics of the NLDE.

\section{Lattice Construction}
\label{sec:LatticeConstruction}

\begin{figure}[b]
\begin{center}
\hspace{0pc} \subfigure{
\label{fig:ex3-a}
\hspace{0in} \includegraphics[scale =.35]{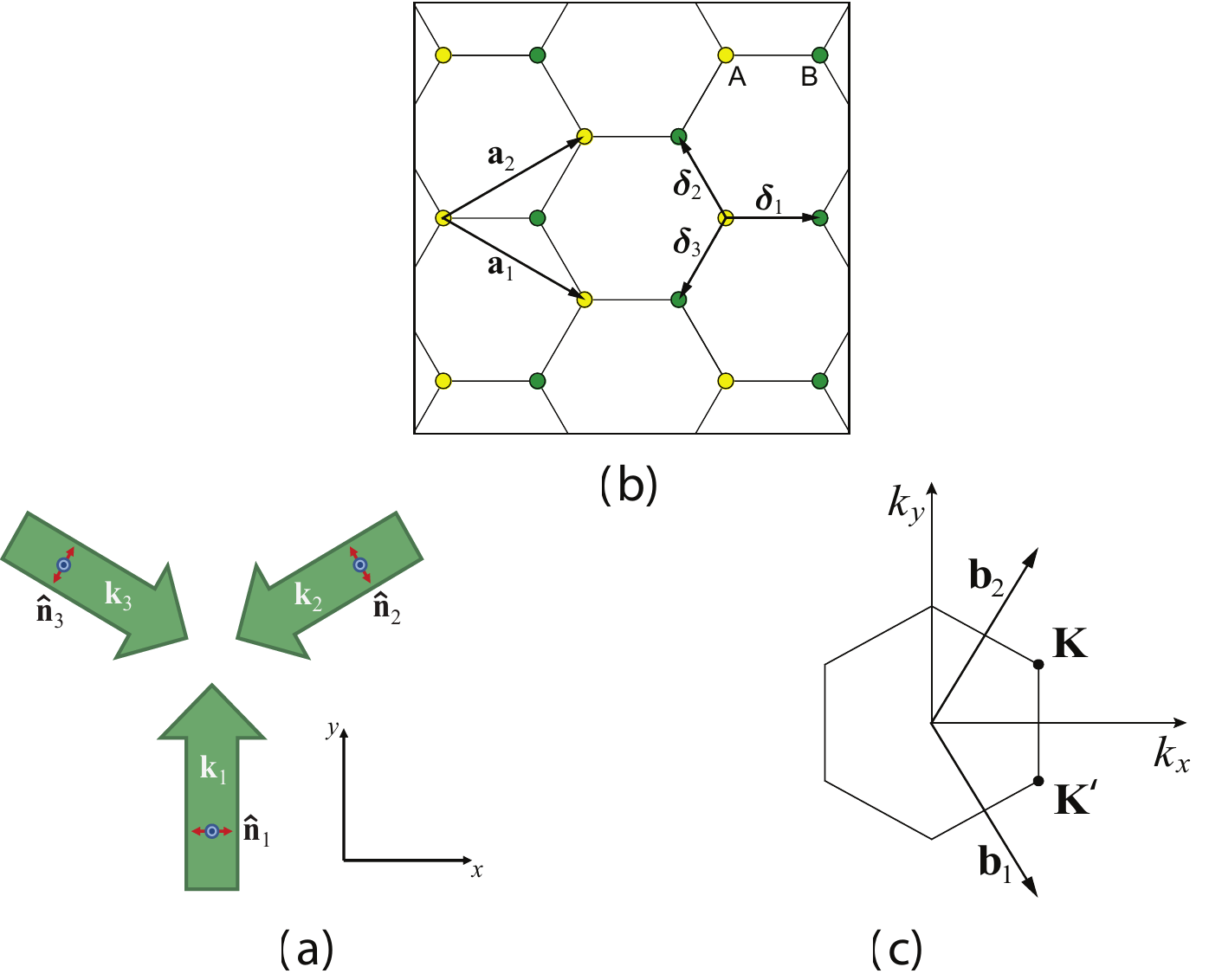}}
\vspace{0pc}
\end{center}
\caption[]{(color online) \emph{The honeycomb optical lattice}. (a) A honeycomb lattice potential can be produced by three co-planar laser beams detuned to the red (blue) of an atomic resonance with polarizations in the plane (orthogonal to the plane). (b) The honeycomb lattice can be described by a hexagonal Bravais lattice with a two-point basis yielding the A and B sublattices. (c) The reciprocal lattice is shown. The single-particle dispersion is linear in the vicinity of two non-equivalent Dirac points at crystal momentum ${\bf K}$ and ${\bf K}'$.}
\label{fig:LattConstruction}
\end{figure}

The honeycomb optical lattice potential is straightforward to implement experimentally~\cite{Salomon93,Sengstock11} using light tuned either to the blue or to the red of an atomic resonance. In both cases, the lattice is formed from three linearly polarized laser beams with co-planar wavevectors separated by an angle of 120$^\circ$, shown in Fig.~\ref{fig:LattConstruction}(a). For a honeycomb lattice formed with blue-detuned light, all three beams have parallel polarizations orthogonal to the plane of propagation. Conversely, the red-detuned lattice has all three laser fields polarized parallel to the plane of propagation. In the latter case, the polarizations make an angle of 120$^\circ$ with respect to one another and the polarization of the net field is spatially dependent. Due to this polarization gradient, the red-detuned optical lattice potential is {\emph{spin-dependent}} as described below.

Optical fields produce an ac Stark shift according to $V $ $=$ $ -\frac{1}{2} \, E_i^{(+)}\,E_i^{(-)}\,\alpha_{i j}$ where ${\bf{E}}^{(\pm)}$ denote the positive/negative frequency components of the optical field and $\alpha_{i j}$ is the dynamic polarizability tensor (which is dependent on the optical frequency). For alkali atoms, the potential can be written as the sum of scalar and vector components $V$ $=$ $ - \frac{1}{2}\,\alpha_{\mathrm{sc}} {\bf{E}}^{(-)}\cdot {\bf{E}}^{(+)} - \frac{1}{2} \, \alpha_{\mathrm{vec}}i \left({\bf{E}}^{(-)} \times {\bf{E}}^{(+)} \right) \cdot {\bf{F}}$, where ${\bf{F}}$ is the total angular momentum operator~\cite{Jessen10}. Here we assume that the detuning of the laser beams from resonance is large in comparison to the hyperfine splitting in the excited state manifolds and neglect a third (tensor) contribution that only becomes significant near resonance. While the scalar light shift is independent of the atom's spin, the vector light shift produces a spin-dependent potential that acts as a spatially dependent effective magnetic field, i.e., $V(\mathbf{r}) = V_{\mathrm{sc}}(\mathbf{r}) + m_F \, g_F \, \mu_B \, B_{\mathrm{eff}}(\mathbf{r})$. Assuming that each of the beams shown in Fig.~\ref{fig:LattConstruction} have equal amplitudes $E_0$, the potential they produce is given by
\begin{eqnarray}
&&\hspace{-1pc} V({\bf{r}}) =   - 2\, V_{\mathrm{sc}} \left\{ 3 + 2\,\hn_1\cdot\hn_2\,\cos[({\bf{k}}_1 - {\bf{k}}_2)\cdot{\bf{r}}]  \right. \nonumber \\
&& \hspace{1pc} \left. + 2\,\hn_2\cdot\hn_3\, \cos[({\bf{k}}_2 - {\bf{k}}_3)\cdot{\bf{r}}] \right.  \nonumber \\
&&\hspace{2pc} \left.  +   2\,\hn_1\cdot\hn_3\, \cos[({\bf{k}}_1 - {\bf{k}}_3)\cdot{\bf{r}}] \right\} \nonumber  \\
&&\hspace{3pc}- 4 \, V_{\mathrm{vec}} \,\left\{ \hn_1 \times \hn_2 \sin[({\bf{k}}_1 - {\bf{k}}_2)\cdot{\bf{r}}]   \right.  \nonumber \\
  &&\hspace{4pc}\left. + \hn_1\times \hn_3\,\sin[({\bf{k}}_1 - {\bf{k}}_3)\cdot{\bf{r}}] \right.  \nonumber \\
 &&\hspace{5pc} \left. + \hn_2 \times \hn_3\,\sin[({\bf{k}}_2 - {\bf{k}}_3)\cdot{\bf{r}}] \right\} \cdot {\bf{F}} \, , 
\label{eqn:LightShiftPot}
\end{eqnarray}
where $\boldsymbol{\hat{\bf{n}}}_i$ are unit vectors denoting the polarization of each beam, $V_{\mathrm{sc}} = \alpha_{\mathrm{sc}} E_0^2/8$, and $V_{\mathrm{vec}} = \alpha_{\mathrm{vec}} E_0^2/8$. In Eq.~(\ref{eqn:LightShiftPot}) we have neglected to include relative phase differences between the beams which only act to translate the lattice in two-dimensions without changing its topology. Note that if the relative phases between the beams vary slowly, the atoms will adiabatically follow the optical lattice potential. The detuning from resonance controls the strength of the vector light shift relative to that of the scalar light shift.

\begin{figure}[]
\begin{center}
 \subfigure{
\label{fig:ex3-a}
\hspace{0in} \includegraphics[width=.35\textwidth]{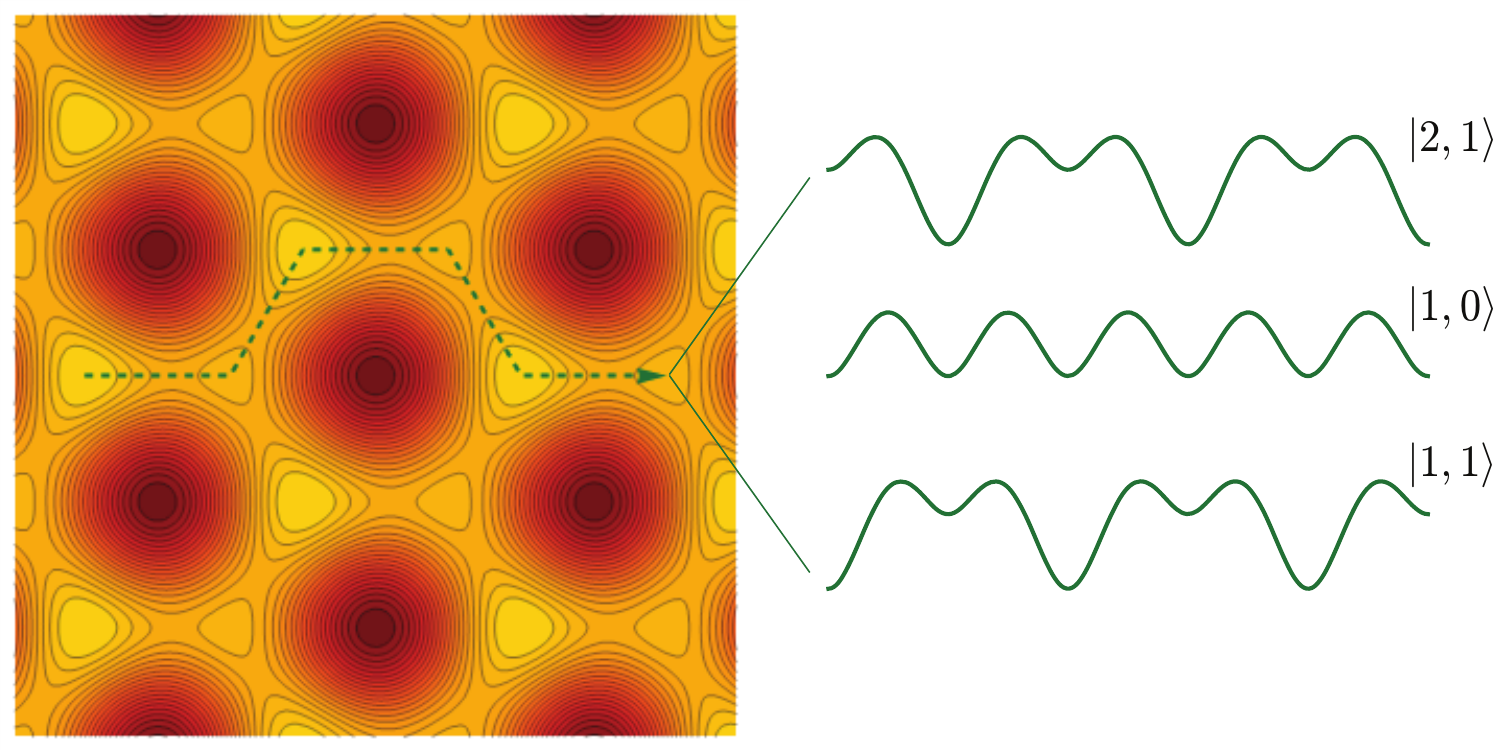}  } 
\vspace{0in}
\end{center}
\caption[]{(color online) \emph{Spin-dependent honeycomb lattice potential.} Honeycomb potential for $^{87}$Rb atoms in state $\left| F, m_F \right\rangle = \left| 2, 1 \right\rangle$ for the case when the wavelength of the lattice light $\lambda_{L} = 422\,{\mathrm{nm}}$. }
\label{SpinDepPotentials}
\end{figure}

The honeycomb lattice produced by the scalar light-shift is described by a hexagonal Bravais lattice with a two-point basis as shown in Fig.~\ref{fig:LattConstruction}(b). In a red-detuned spin-dependent lattice, the depths of the A and B sublattices can be asymmetric, e.g., $\left|F, m_F \right\rangle = \left|2, 1  \right\rangle$ or $\left|1, 1  \right\rangle$, or symmetric, e.g., $\left|F, m_F \right\rangle = \left|1, 0  \right\rangle$, depending on the internal state of the atom. An A/B sublattice asymmetry produces a mass gap at the Dirac points. For a red-detuned lattice with polarizations in the plane, the mass gap $2 |m_s| \approx 7 m_F V_\mathrm{vec}$ separates the s-bands of the A and B sublattices at the Dirac point. Figure~\ref{SpinDepPotentials} shows the optical potential produced for $^{87}$Rb atoms in different hyperfine states when the lattice is formed from $\lambda_L = 422\,{\mathrm{nm}}$ light red detuned from the $5S - 6P$ transition~\cite{Sengstock11}.

\section{Preparing a BEC at a Dirac point}
\label{sec:DiracPointPrep}

Study of the NLDE will require that the BEC be prepared at a Dirac point, i.e.,  $\mathbf{K}$ or $\mathbf{K'}$ in Fig.~\ref{fig:LattConstruction}(c). Several experimental methods can potentially accomplish this: first, loading a BEC into the lowest-energy Bloch state and subsequently applying a constant acceleration for a fixed duration; second, loading an initially stationary BEC directly into a Bloch state at a Dirac point $\mathbf{K}$ by adiabatically applying a moving lattice potential which maintains a constant velocity $\hbar \mathbf{K}/M$; and third, loading a BEC into the lowest-energy Bloch state and subsequently populating a Dirac point by Bragg scattering using auxiliary fields. The first two methods have potential deficiencies.  With regard to the first method, a dynamical instability may exist for intermediate values of the crystal momenta as it linearly increases from $\mathbf{0}$ to $\mathbf{K}$~\cite{StamperKurn1999}.  For the second method, the timescale required for adiabaticity is divergent since there is no gap for crystal momenta along the Brillouin zone boundary in the absence of a lattice potential.  Hence, we consider here the method of populating a Dirac point by inducing Bragg scattering between crystal momenta $\mathbf{0}$ and $\mathbf{K}$ using auxiliary laser fields.

It is straightforward to populate the lowest-energy Bloch state of a honeycomb lattice by adiabatically increasing the lattice depth as demonstrated in Ref.~\cite{Sengstock11} where both the BEC and the lattice are stationary in the lab frame.  Here we will assume that the BEC is in a hyperfine state with $m_F \neq 0$ and a spin-dependent potential is used. This is so that only the sublattice with the lowest energy, assumed here to be the A sublattice, becomes occupied~\cite{Sengstock11}. Starting from this initial condition, Bragg scattering to a Bloch state at a Dirac point can be accomplished by applying two laser fields with wavevectors $\mathbf{k}_{\mathrm{b1}}$ and $\mathbf{k}_{\mathrm{b2}}$, which satisfy $\mathbf{k}_{\mathrm{b1}} - \mathbf{k}_{\mathrm{b2}} = \mathbf{K}$ and have frequencies $\omega_{\mathrm{b1}}$ and $\omega_{\mathrm{b2}}$ with the condition that $\omega_{\mathrm{b1}} - \omega_{\mathrm{b2}} = \Delta \omega = [E_A({\mathbf{K}}) - E_A({\mathbf{0}})]/\hbar$. In this expression, the function $E_A$ gives the dispersion relation for the lower band of a honeycomb lattice with A/B sublattice asymmetry, which corresponds approximately to full occupation of the A sublattice. Hence, throughout our analysis we will designate the lower band using the subscript A. These fields produce a Stark shift potential
\begin{eqnarray}
V_{\mathrm{Bragg}}(\mathbf{r}) = \frac{1}{2}  V_B \left[ \cos(\mathbf{K} \cdot \mathbf{r} - \Delta \omega \, t) + 1\right], 
\end{eqnarray}
where $V_B$ sets the strength of the potential. This potential couples the Bloch wavefunctions $\psi_{A, \mathbf{K}}(\mathbf{r}) =  e^{i \mathbf{K} \cdot \mathbf{r}} \, u_{A, \mathbf{K}} (\mathbf{r})$ and $ \psi_{A, \mathbf{0}} (\mathbf{r}) =  u_{A, \mathbf{0}} (\mathbf{r})$ where
$u_{A, \mathbf{K}} (\mathbf{r})$ and $u_{A, \mathbf{0}} (\mathbf{r})$ have the same periodicity as the lattice.  Thus, both functions can be written in the form
\begin{eqnarray}
u_{A, \mathbf{K}}(\mathbf{r}) = \sum_{\mathbf{Q}} \, C^{A, \mathbf{K}}_{\mathbf{Q}} \, e^{i \, \mathbf{Q} \cdot \mathbf{r}},
\end{eqnarray}
where the sum over $\mathbf{Q}$ includes all vectors in the reciprocal lattice space. The coefficients $C^{A, \mathbf{K}}_{\mathbf{Q}}$ can be calculated for a honeycomb lattice of arbitrary scalar and vector potential depths, $V_{\mathrm{sc}}$ and $V_{\mathrm{vec}}$ respectively, by numerically computing the band structure for the potential given in Eq.~(\ref{eqn:LightShiftPot})~\cite{Lee:2009}.

\begin{figure}[]
\begin{center}
 \subfigure{
\label{fig:ex3-a}
\hspace{-.2in} \includegraphics[width=.375\textwidth]{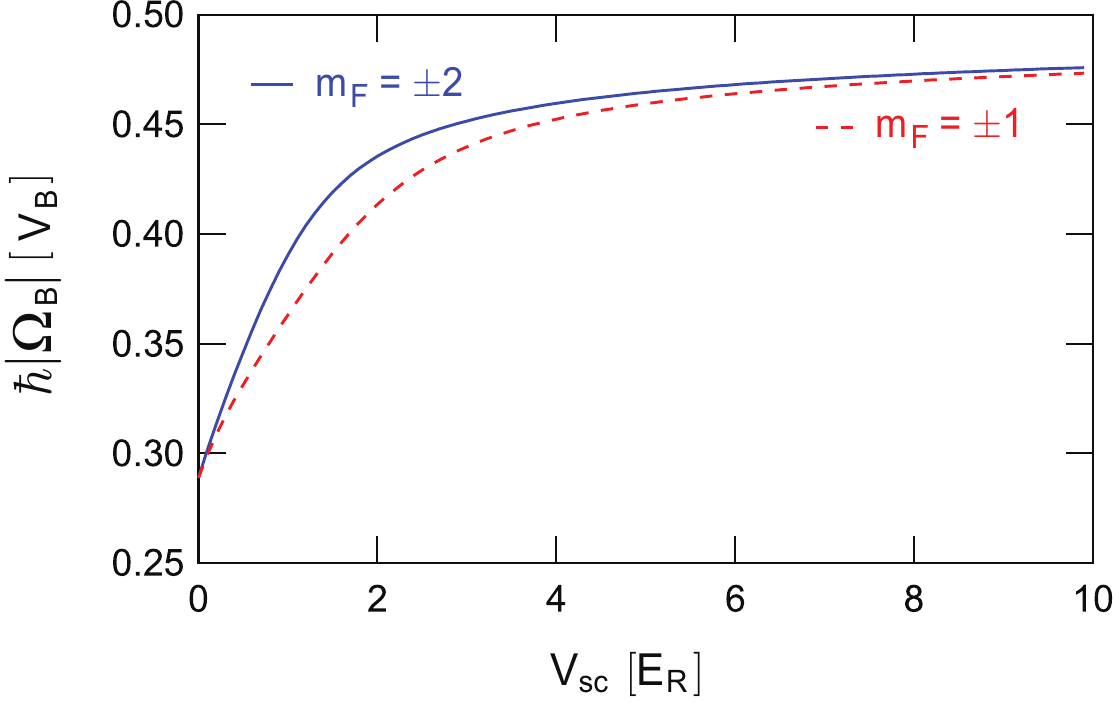}  } 
\vspace{0in}
\end{center}
\caption[]{(color online) \emph{Bragg scattering in a spin-dependent honeycomb lattice.} Rabi frequencies for transitions between $\psi_{A,{\bf{0}}}$ and $\psi_{A,{\bf{K}}}$ when $^{87}$Rb atoms are in the $m_F = \pm 2$ (solid blue) and $\pm 1$ (dashed red) states. Here, we assume that $V_\mathrm{vec}/V_\mathrm{sc} = 0.13$.}
\label{BraggScatt0toK}
\end{figure}

Application of the Bragg scattering potential then results in Rabi oscillation between $\psi_{A,\mathbf{0}}$ and $\psi_{A,\mathbf{K}}$ with a Rabi frequency $\Omega_B$ given by
\begin{eqnarray}
\Omega_B = \frac{V_B}{2\, \hbar} \, \sum_{\mathbf{Q}} (C^{A, \mathbf{K}}_{\mathbf{Q}})^* \, C^{A, \mathbf{0}}_{\mathbf{Q}}.
\end{eqnarray}
Figure~\ref{BraggScatt0toK} shows numerical calculations for $\left| \hbar\, \Omega_B \right|$ as a function of the depth of the honeycomb lattice $V_{\mathrm{sc}}$ in units of the depth of the Bragg scattering lattice $V_{B}$.  For these calculations, we assume that $V_\mathrm{vec}/V_\mathrm{sc} = 0.13$ which can be achieved with $^{87}$Rb using 422 nm light which is red-detuned from the 5S - 6P transition. The entire population of atoms in state $\psi_{A,\mathbf{0}}$ can be transferred to $\psi_{A,\mathbf{K}}$ by applying the Bragg scattering potential for a time $\tau_\pi = \pi/\Omega_B$ provided that $V_B$ is chosen such that $\hbar/\tau_\pi$ is significantly smaller than the energy splitting between bands.

A particularly useful feature of using a honeycomb lattice potential with A/B sublattice asymmetry for preparation is that both the initial and final Bloch states ($\psi_{A, \mathbf{0}}$ and $\psi_{A, \mathbf{K}}$) have a group velocity relative to the lattice equal to zero. If the lattice is stationary with respect to the lab frame, the condensate will then also be stationary both before and after transfer to the Dirac point. Note that the condensate would not remain stationary if it were transferred to the Dirac point by Bragg scattering in a lattice with A/B sublattice symmetry (i.e. no mass gap). In this case, the lower and upper $s$-bands are degenerate at the Dirac point and the eigenstates can be chosen from a two-dimensional subspace of degenerate states spanned by two Bloch wavefunctions. Application of the $V_{\mathrm{Bragg}}$ potential breaks this degeneracy and excites the eigenstate which moves in the same direction as that of the walking standing wave potential $V_{\mathrm{Bragg}}$. This particular eigenstate has a group velocity magnitude equal to $c_l$ in the frame of the lattice. The orthogonal eigenstate has a group velocity with the same magnitude but in the opposite direction and is not coupled by $V_{\mathrm{Bragg}}$ to the Bloch state with zero crystal momentum.

Once the condensate has been prepared at a Dirac point in a lattice with A/B sublattice asymmetry by Bragg scattering, the atoms can be transferred to a hyperfine state that does not experience the vector light shift and therefore no mass gap, e.g., $\left| F, m_F \right\rangle = \left| 1, 1 \right\rangle \rightarrow \left|1, 0 \right\rangle$, using a radio-frequency (rf) or microwave (mw) field. For a spatially homogeneous rf/mw field, the transition matrix element is proportional to the spatial overlap of the initial and final spatial wavefunctions which are not orthogonal since they experience different lattice potentials. A spatially homogeneous rf/mw field cannot change the crystal momentum which is therefore conserved in the transition.

In the absence of a vector light shift, the A and B sublattices are symmetric and there is no mass gap, yielding two degenerate Bloch states at the Dirac point $\mathbf{K}$. Two orthogonal basis states that span the degenerate subspace of eigenstates can be chosen to be states which have probability current density $\mathbf{j} \equiv - i \frac{\hbar}{2 M} \left( \Psi^* \nabla \Psi - \Psi \nabla \Psi^*\right) = \mathbf{0}$ but are respectively localized on either the A or B sublattice sites.  A state prepared at the Dirac point of a lattice with a mass gap will have significant spatial overlap with one of these basis states and vanishing overlap with the orthogonal state.  For example, for parameters identical to those realized in ~\cite{Sengstock11}, i.e., $V_{\mathrm{sc}} = 4\, E_R$ and $V_{\mathrm{vec}}/V_{\mathrm{sc}} = 0.065$, the magnitude of the inner product between the initial and final states for wavefunctions localized on the same sublattice is $\left| \left\langle A, \mathbf{K}, m_F = 1 \right| \left. A, \mathbf{K}, m_F = 0 \right\rangle \right| = 0.995$ whereas $\left| \left\langle A, \mathbf{K}, m_F = 1 \right| \left. B, \mathbf{K}, m_F = 0 \right\rangle \right| = 0$. Thus, by driving a transition between internal states with a rf/mw field, a condensate which remains stationary can be prepared at the Dirac point of a honeycomb lattice with no mass gap. The state produced will only have amplitude in sites of the A sublattice.  In the next section we discuss how the condensate can be coherently transferred between A and B sublattices by modulating the lattice potential.

\section{Coherent Transfer Between Sublattices}
\label{sec:SublatticeTransfer}

 As previously discussed, when $m_F \neq 0$ the lattice has an A/B sublattice asymmetry which produces a mass gap $2 \left| m_s \right|$ separating the $s$-bands of the A and B sublattices at the Dirac point. Note that in the fully covariant NLDE, the mass gap will appear as a factor of $m_s c_l$ multiplying the spinor wavefunction, where $c_l$ is the effective speed of light. In such cases when $m_F \neq 0$, transitions between Bloch states $\psi_{A,\mathbf{K}}$ and $\psi_{B,\mathbf{K}}$ can be driven by applying a periodic perturbation $H_m(\bf{r}) \cos\omega_s t$ where $\hbar \omega_s = 2 \left| m_s \right| $, and $H_m(\mathbf{r})$ is chosen to exclusively couple pairs of Wannier states $w_A$ and $w_B$ localized on adjacent A and B sites of a given unit cell, e.g., $\left\langle w_A({\bf{r}} - {\bf{r}}_A) \right| H_m \left| w_B({\bf{r}} - {\bf{r}}_A' - {\boldsymbol{\delta}}_1)\right\rangle = \hbar \, \Omega_{m} \, \delta_{{\bf{r}}_A, {\bf{r}}_A'}$ where ${\boldsymbol{\delta}}_1$ is the displacement between an A site and {\emph{one}} of its three neighboring B sites.  A perturbation which only couples pairs of Wannier states separated by one of the nearest neighbor displacement vectors, e.g., ${\boldsymbol{\delta}}_1$,  conserves the crystal momentum so that $\left\langle \psi_{A,\bf{K}+\bf{q}} \right| H_m \left| \psi_{B,{\bf{K}}+{\bf{q}}'} \right\rangle = e^{i ({\bf{K}} + {\bf{q}}) \cdot {\boldsymbol{\delta}}_1} \, \Omega_m \, \delta_{{\bf{q}},{\bf{q}}'}$. A suitable perturbation $H_m$ can be experimentally realized by modulating the amplitude of one of the lattice laser fields, which provides an anisotropic modulation of the tunneling matrix elements that discriminates tunneling in one direction, while simultaneously frequency modulating the other two fields, which periodically shakes the lattice along the same direction. Amplitude modulation of the field ${\bf{E}}_1$  and frequency modulation of ${\bf{E}}_2$ and ${\bf{E}}_3$ in Fig.~\ref{fig:LattConstruction}, for example, yields a periodic perturbation with a spatial dependence given by
\begin{widetext}
\begin{eqnarray}
H_m(\mathbf{r}) = V_m \left[ \cos({\bf{k}}_1 - {\bf{k}}_2)\cdot{\bf{r}} + \cos({\bf{k}}_1 - {\bf{k}}_3)\cdot{\bf{r}} + \sqrt{3} \, \frac{V_{\mathrm{vec}}}{V_{\mathrm{sc}}}  m_F  \left \{ \sin({\bf{k}}_1 - {\bf{k}}_2)\cdot{\bf{r}} + \sin({\bf{k}}_1 - {\bf{k}}_3)\cdot{\bf{r}} \right\} - \kappa \, \boldsymbol{\hat{\boldsymbol{\delta}}}_1 \cdot \bf{r} \right],\; 
\label{eqn:ABTransferPerturb}
\end{eqnarray}
\end{widetext}
where $\kappa$ depends on the relative amplitudes of the perturbations.  The last term in the square brackets describes shaking of the lattice along the ${\boldsymbol{\delta}}_1$ direction while the other terms act to anisotropically modulate the tunneling matrix elements between nearest neighbors with tunneling in the ${\boldsymbol{\delta}}_1$ direction distinguished from the other two.

The perturbations resulting from amplitude and frequency modulation both anisotropically couple a Wannier state $w_A$ to Wannier states $w_B$ localized on the three neighboring sites, but discriminate tunneling in the ${\boldsymbol{\delta}}_1$ direction with different relative strengths. By adjusting the relative amplitude of the two perturbations, nearest neighbors in the ${\boldsymbol{\delta}}_1$ direction can be strongly coupled with negligible coupling to neighboring sites in the other two directions.

\begin{figure}[]
\includegraphics[width=.2\textwidth]{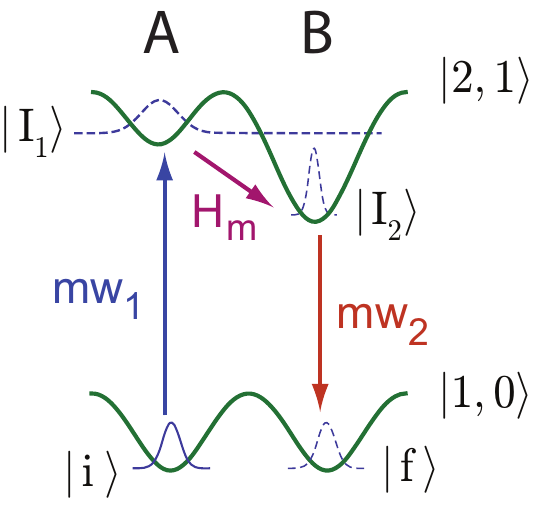}
\caption{(color online) \emph{Coherent transfer between sublattices A and B}. Three step process of exciting atoms from the A sublattice with hyperfine state $|1, 0\rangle$ (no sublattice asymmetry) to the hyperfine state $|2,1\rangle$ via the rf/mw transition $\mathrm{mw}_1$, then from the A sublattice to the B sublattice via the perturbation $\mathrm{H}_\mathrm{m}$, and finally back to the $|1, 0\rangle$ hyperfine state via the $\mathrm{mw}_2$ transition ($\mathrm{mw} = \mathrm{microwave}$).}
\label{fig:SublatticeTransfer}
\end{figure}

To coherently transfer a condensate between sublattices when the condensate is initially in an internal state with $m_F = 0$, which does not experience an A/B sublattice asymmetry, an rf/mw transition can be applied to couple to an intermediate internal state with $m_F \neq 0$ that does experience an A/B sublattice asymmetry. Modulation of the lattice potential with the perturbation $H_m({\bf{r}}) \, \cos \omega_m t$ can then be applied to drive transitions between the A and B sublattices as described above provided that $\hbar \omega_m$ equals the mass gap for the condensate with $m_F \neq 0$. The atoms can be subsequently transferred back to the original internal state via an ensuing rf/mw transition. A suitable transition sequence for $^{87}$Rb atoms in a spin-dependent lattice is depicted in Fig.~\ref{fig:SublatticeTransfer}. Assuming that the rf/mw field is homogeneous over the size of the sample, the crystal momentum is conserved in this process though the sublattice index is changed.

\section{Coherent Transfer Between Dirac Points by Bragg Scattering}
\label{sec:BraggScatter}

Once a BEC has been prepared at a Dirac point ${\mathbf{K}}$, coherent transfer to the non-equivalent Dirac point ${\mathbf{K'}}$ can be accomplished by Bragg scattering from a lattice formed using auxiliary laser fields~\cite{Sengstock10}. In this case, the two additional laser fields have wavevectors $\mathbf{k}_{\mathrm{b1}}$ and $\mathbf{k}_{\mathrm{b2}}$ where $\mathbf{k}_{\mathrm{b1}} - \mathbf{k}_{\mathrm{b2}} = \mathbf{K'} - \mathbf{K} = - k_L\,\hy$ in the frame of the lattice. The lattice produced by these fields couples a BEC at crystal momentum ${\mathbf{K}} = k_L \left( \sqrt{3} \, \hx /2+   \hy/2 \right)$ to a BEC with crystal momentum ${\mathbf{K'}} = k_L  \left( \sqrt{3} \, \hx /2-   \hy/2 \right)$ by Bragg scattering. Since the energies of the two coupled Dirac points are identical, resonance occurs when the optical frequencies of the auxiliary fields are equal and the standing wave they form is stationary in the frame of the honeycomb lattice.

\begin{figure}[h]
\begin{center}
 \subfigure{
\label{fig:ex3-a}
\hspace{-.2in} \includegraphics[width=.375\textwidth]{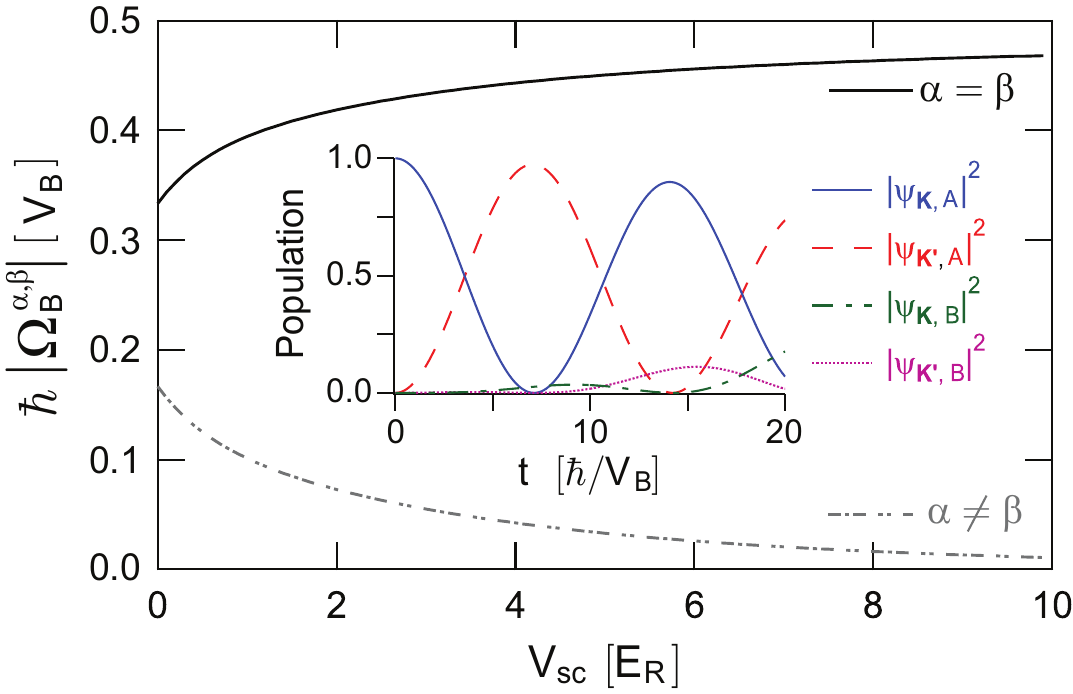}  } 
\vspace{0in}
\end{center}
\caption[]{(color online) \emph{Bragg scattering between Dirac points.} Rabi frequency for transitions between non-equivalent Dirac points for cases where the sub-lattice index remains the same (solid blue) or changes (dashed red) as functions of the depth of the scalar part $V_{\mathrm{sc}}$ of the optical lattice potential. (Inset) Time dependence of the sublattice populations at the Dirac points $\bf{K}$ and $\bf{K}'$ for an optical lattice depth of $V_{\mathrm{sc}} = 4 E_R$.  }
\label{BraggScattMatElem}
\end{figure}

In the frame of the lattice, the applied potential $V_{\mathrm{Bragg}}(\mathbf{r}) = (1/2) V_B \, \left[ \cos(\mathbf{K} - \mathbf{K}')\cdot \mathbf{r} + 1 \right]$. This potential couples the degenerate Bloch wavefunctions $\psi_{\mathbf{K}}(\mathbf{r})$ and $\psi_{\mathbf{K'}} (\mathbf{r})$. The matrix element coupling $\psi_{\mathbf{K}}$ and $\psi_{\mathbf{K'}}$ is then given by
\begin{eqnarray}
\Omega_B^{\alpha,\beta} = \frac{V_B}{2 \, \hbar} \, \sum_{\mathbf{Q}} \left(C_{\mathbf{Q}}^{\alpha,\mathbf{K}'} \right)^* \, C^{\beta,\mathbf{K}}_{\mathbf{Q}},
\end{eqnarray}
where the coefficients $C_{\mathbf{Q}}^{\alpha,\mathbf{K}}$ are identical to those defined in Sect.~\ref{sec:DiracPointPrep} where the index $\alpha$ designates the sublattice on which the condensate is localized. These coefficients can be found by numerically computing the band structure for the potential given in Eq.~(\ref{eqn:LightShiftPot})~\cite{Lee:2009}. In this case of a condensate in an internal state with $m_F = 0$ which does not have a gap at the Dirac points, there are four degenerate Bloch wavefunctions corresponding to the two possible inequivalent Dirac points (${\bf{K}}$ and ${\bf{K}}'$) and the two possible sublattices (A and B). In the tight-binding limit, i.e. $V_{\mathrm{sc}} \gg E_R$, the Bragg scattering lattice only couples Bloch states at the non-equivalent Dirac points that are localized on the same sublattice.  In this limit, application of the Bragg scattering lattice will induce Rabi oscillations with frequency $\Omega_{B}^{\alpha,\alpha}$ between condensates localized on the same sublattice but at the non-equivalent Dirac points. For shallower depths of the honeycomb lattice, all four degenerate Bloch states will be coupled and the dynamics will be more complicated. However, even for a moderate lattice depth $V_{\mathrm{sc}} = 4 \, E_R$, the coupling between different sublattices is small enough that the dynamics are nearly identical to those of two coupled Bloch states. Starting from a BEC initially prepared at a single Dirac point $\mathbf{K}$, application of the Bragg scattering potential will cause the amplitude to Rabi oscillate between $\psi_{\mathbf{K}}$ and $\psi_{\mathbf{K'}}$ with a Rabi oscillation frequency $\Omega_{\mathrm{Bragg}} = 2\, \left| \left\langle \mathbf{K'} \left| V_{\mathrm{Bragg}} \right| \mathbf{K} \right\rangle \right|$.  The pulse duration of the auxiliary fields can be controlled to produce an arbitrary superposition of BECs at ${\mathbf{K}}$ and ${\mathbf{K'}}$ -- with a $\pi/2$-pulse $\tau_{\pi/2} = (\pi/2)/\Omega_{\mathrm{Bragg}}$ producing an equal superposition. This process is depicted in Fig.~\ref{BraggScattMatElem}, where we have plotted the Rabi frequency versus the depth of the scalar part of the optical lattice potential.

\section{Vortex solutions and linear stability analysis}
\label{sec:VortexSolutions}

\begin{figure}[]
\hspace{-.5pc} \includegraphics[width=.45\textwidth]{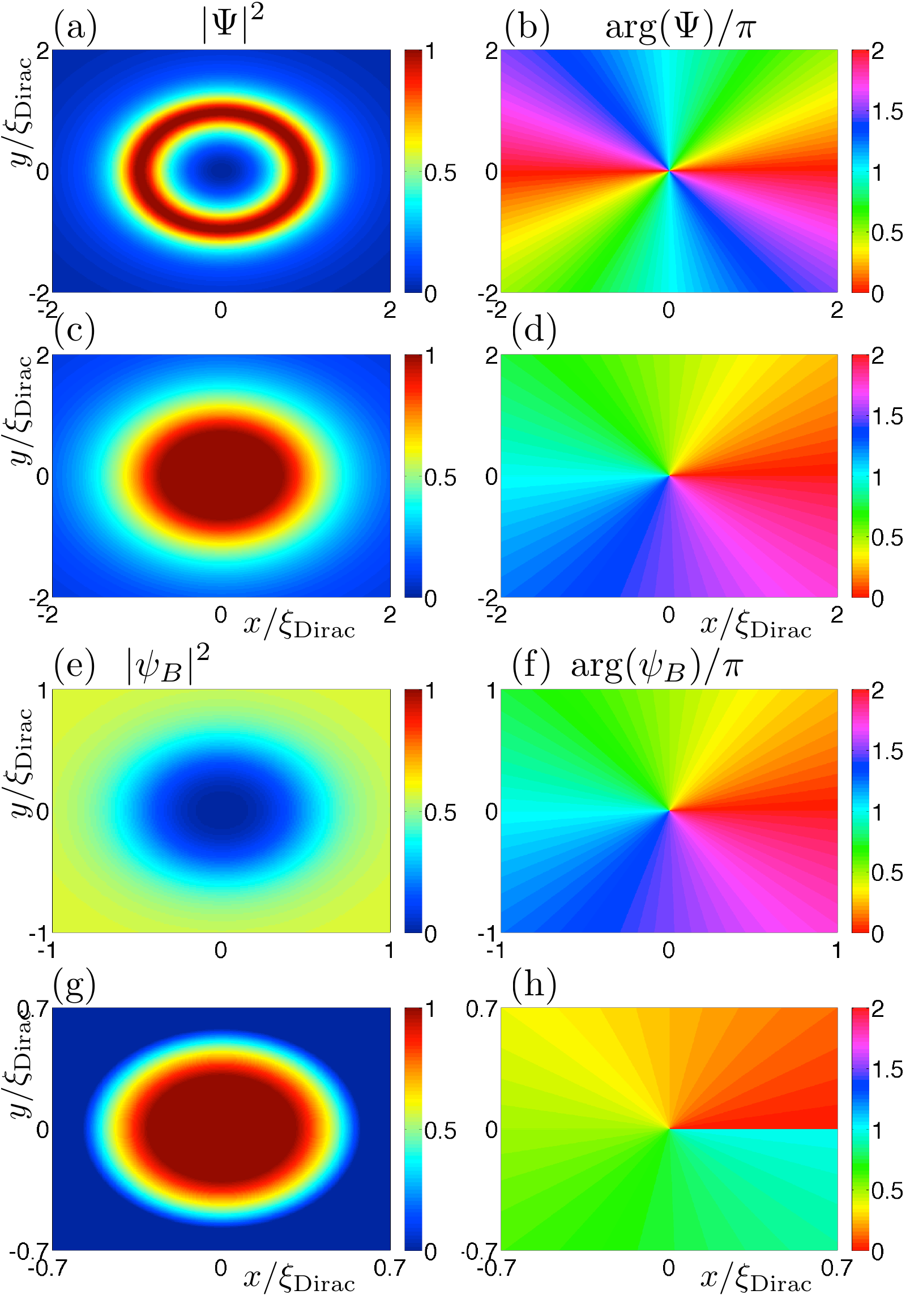}
\caption[]{(color online) \emph{Plots of relativistic vortices.} Total density and phase of (a,b) $\ell =2$ ring-vortex, (c,d) B sublattice of Mermin-Ho skyrmion, (e,f) ring-vortex/soliton, (g,h) half-quantum vortex, or semion. All these vortices and more can be made by variations on the experimental techniques of Figs.~\ref{BraggScatt0toK}-\ref{fig:SublatticeTransfer}.}
\label{RelativisticVortices}
\end{figure}

We analytically and numerically obtain seven physically distinct NLDE vortex types as follows. (i) The vortex/soliton is a bright soliton or density peak in the center in the first component with a vortex of phase winding $2\pi$ around the outside in the second. (ii) The ring-vortex/soliton is also a bright soliton in the first component, but the vortex component is a ring peaked near the healing length $r = \xi_\mathrm{Dirac}$. (iii) The Anderson-Toulouse skyrmion has the same core structure as the vortex/soliton, but the spinor components are continuously interchanged as the distance from the core increases, while staying within the bounds $|\psi_A|, \, |\psi_B| \in (0, 1)$ and conserving total density $|\psi_A|^2 + |\psi_B|^2 = 1$. (iv) The Mermin-Ho skyrmion again has similar behavior near the core but the soliton (vortex) amplitude decreases (increases) monotonically away from the core  within the bounds $\mathrm{cos}(\pi/4) < \psi_A < 1$ and $0 < \psi_B <  \mathrm{cos}(\pi/4)$. (v) The half-quantum vortex or semion is characterized by a phase discontinuity such that far from the core the amplitudes have the form $\psi_A \propto \mathrm{cos}(\theta/2)$ and $\psi_B \propto \mathrm{sin}(\theta/2)$; the additional $\pi$ phase is accounted for by a rotation between the Dirac spinor components. So far, all of these solutions have one unit of angular momentum, $\ell =1$, either a phase winding of $2\pi$ in one component or a winding of $\pi$ in each component. Additionally, for arbitrary phase winding ($\ell > 1$ with $\ell \in \mathbb{N}$) (vi) ring-vortices and (vii) topological vortices exist with $\ell- 1$ ($\ell$) units of winding in the first (second) spinor component, but differ in their asymptotic form. Component amplitudes for the ring-vortex peak at around one healing length from the core and quickly decay for large $r$. On the other hand, topological vortices retain non-zero density far from the core. Several representative vortices are plotted in Fig.~\ref{RelativisticVortices}. In addition, Table~\ref{table2} details the functional form of each vortex type. We note the similarities to realizations of skyrmions in a spin-2 BEC~\cite{Leslie2009}. All of the vortices here can be created using straightforward variations of the transition sequence depicted in Fig.~\ref{fig:SublatticeTransfer}, as we discuss in detail in Sec.~\ref{sec:CreateVortex}. 
\begin{table*}[phtb]
\begin{center}
\begin{tabular}{  c   c  c   p{2.3cm}  }
\hline   \hline
{\bf Vortex type } & {\bf Winding }  & { \bf Analytic form of} $\Psi({\bf r})$   &  {\bf Topology}    \\
 \hline 
 Vortex/soliton & $\ell =1$  &$  \left[  i \frac{ 1 }{ \sqrt{1 + \, ( r/r_0)^2} } , \, e^{  i \theta}\!  \frac{ (r/ r_0)}{ \sqrt{1 + \, ( r/r_0)^2} } \right]^T $   & $| \psi_A(\infty)| = 1$   \\
Ring-vortex/soliton &$ \ell = 1$  & $  \left[  i  \frac{ 1}{ \sqrt{1 + \, ( r/r_0)^4} } , \,  e^{i \theta}\! \frac{ (r/ r_0)}{ \sqrt{1 + \, ( r/r_0)^4} } \right]^T $      &   non-topological  \\
Anderson-Toulouse skyrmion &$\ell =1$  & $  \left[  i \,   \mathrm{cos} \varphi (r/r_0) , \, e^{ i \theta}  \mathrm{sin} \varphi (r/r_0)  \right]^T $ &  $\varphi (\infty) = 0$      \\
 Mermin-Ho skyrmion & $\ell = 1$ &$  \left[  i \,  \mathrm{cos} \varphi (r/r_0) , \, e^{ i \theta}  \mathrm{sin} \varphi (r/r_0)  \right]^T $  &  $\varphi (\infty) = \pi /4 $     \\
Half-quantum vortex & $\ell =1$  & $  \left[  i  \mathrm{cos}\,  \theta/2 , \;  \mathrm{sin} \, \theta /2  \right]^T $  & $ |\Psi (\infty)| = 1$  \\ 
Ring-vortex & $\ell =  2, 3, 4,  ...$ & $  \left[  i e^{ i (\ell -1)  \theta}\! \frac{ (r/ r_0)^{ \ell - 1 } }{ \sqrt{1 + \, ( r/r_0)^{8(\ell -1/2)}} } , \,e^{ i  \ell  \theta}\!  \frac{ (r/r_0)^{3 \ell -2 } }{ \sqrt{1 + \, ( r/r_0)^{8(\ell -1/2)}} } \right]^T $  & non-topological  \\
General topological vortex & $\ell =  2, 3, 4,  ...$&  Numerical  shooting method & $| \psi_A(\infty)| = 1$ \\
\hline \hline
\end{tabular}   
{\caption{\emph{Vortex solutions of the NLDE.} Solutions are described by their phase winding, closed-form expression, and topological properties. Solutions which retain non-zero density far from the core have an associated conserved topological charge, and we state their asymptotic form. Note that $r_0$ is the length scale associated with the chemical potential or the interaction strength depending on the particular solution.}   \label{table2}}
\end{center}
\end{table*}

We elaborate here on the methods used to obtain vortex solutions of the NLDE. The NLDE treats the entire Dirac four-spinor. In its simplest realization without mass gaps and in tight binding the upper two components, called a \textit{Weyl spinor}, are decoupled from the lower two, and can be written $\Psi=(\psi_A, \psi_B)^T$. We obtain vortex solutions by expressing the spinor components in the form: $\psi_A(r, \theta, t) = \pm i \,  f_A(r) e^{i (\ell -1) \theta} \, e^{-i \mu t/\hbar}$, $\psi_B(r, \theta, t) =  f_B(r) e^{i \ell  \theta} \, e^{-i \mu t/\hbar}$, and writing the NLDE in plane-polar coordinates:
\begin{eqnarray}
\hspace{-1.2pc} - \hbar c_l \! \left(\! \partial_{r } +  \frac{\ell}{r} \right)\! f_B(r) + U \left|f_A(r)\right|^2\! f_A(r)   \!  &=&\!  \mu  f_A(r) \label{eqn:CondPsi7} \\
\hspace{-1.2pc}   \hbar c_l\! \left( \!\partial_{r }+ \frac{1\! - \! \ell }{r} \right)\! f_A(r) + U  \left|f_B(r)\right|^2 \! f_B(r)\!   &=& \!   \mu f_B(r)  \label{eqn:CondPsi8} , 
\end{eqnarray}
where $\ell$ is the integer phase winding and the other parameters are defined in Table~\ref{table1}. For the case $\mu=0$, Eqs.~(\ref{eqn:CondPsi7})-(\ref{eqn:CondPsi8}) give closed form expressions for the radial amplitudes $f_A$ and $f_B$. These are the ring-vortex/soliton ($\ell =1$) and general ring-vortex ($\ell >1$) solutions. For the case $\mu \ne 0$, closed form solutions exist in some cases while others are obtained using a numerical shooting method (see Table~\ref{table2}).

\begin{figure}[h]
\centering
\hspace{-.2in}\subfigure{
\label{fig:ex3-b}
\includegraphics[width=.5\textwidth]{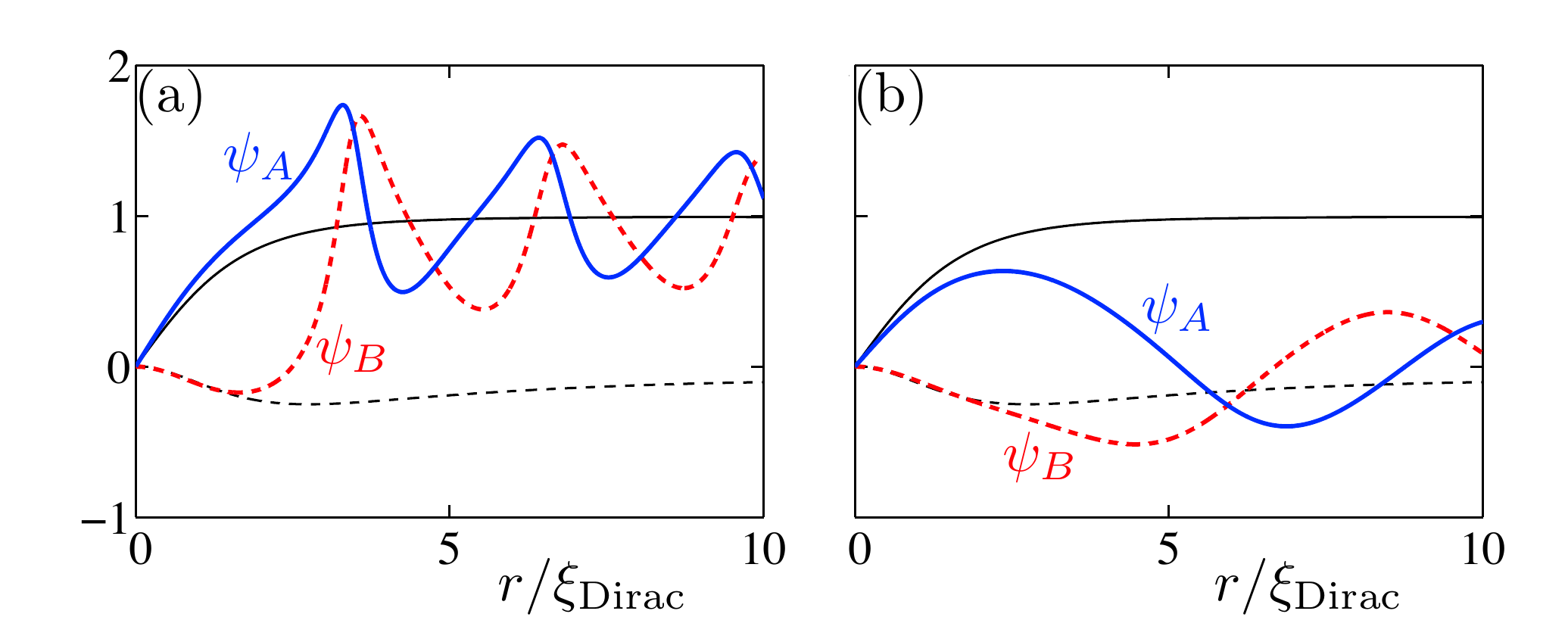}}\\
\caption[]{(color online) \emph{Vortex radial profiles}. Numerical shooting for $\ell=2$ vortex. (a) For $a_1 > a_1^{\textrm{vortex}}$, the solution overshoots to an excited state of the vortex. (b) For $a_1 <  a_1^{\textrm{vortex}}$, the solution undershoots and converges to the linear solution Bessel functions. Note that the solid blue and dashed red plots are the A and B sublattice radial wavefunctions, respectively. The solid black and dashed black plots are the exact solutions for the A and B sublattice radial wavefunctions, respectively.}
\label{NumVortexConv}
\end{figure}

Numerical solutions for general values of the chemical potential $\mu$ and arbitrary winding $\ell$ were obtained by the method of numerical shooting~\cite{CarCla06}. We express Eqs.~(\ref{eqn:CondPsi7})-(\ref{eqn:CondPsi8}) in terms of the dimensionless radial variable $\chi \equiv r/\xi_\mathrm{Dirac}$, where $\xi_\mathrm{Dirac} = \hbar c_l/U$ is the quasi-two-dimensional renormalized healing length discussed in Table~\ref{table1}. The functions $f_A(\chi)$ and $f_B(\chi)$ are then expanded in a power series around $\chi=0$
\begin{eqnarray}
f_A(\chi) = \sum_{j=0}^\infty a_j \chi^j \, , \;\;\;\;  f_B(\chi) = \sum_{j=0}^\infty b_j \chi^j \, , \label{expansions}
\end{eqnarray}
where the $a_j$ and $b_j$ are the expansion coefficients. Since we are solving two coupled first order equations, we require the initial conditions $f_A(0)$ and $f_B(0)$. Substituting into Eqs.~(\ref{eqn:CondPsi7})-(\ref{eqn:CondPsi8}) gives us the core behavior: 
\begin{eqnarray}
f_A(0) \sim \chi^{\ell-1} \; , \;\;\;\;\; f_B(0) \sim \chi^\ell\, . \label{corevalues}
\end{eqnarray}
These core values indicate that the first nonzero coefficients for a given choice of $\ell$ are $a_{\ell-1}$ and $b_\ell$, where $a_{\ell-1}$ is sufficient to determine all other coefficients for both expansions in Eq.~(\ref{expansions}). Equations~(\ref{eqn:CondPsi7})-(\ref{eqn:CondPsi8}) are then discretized using either a finite difference or fourth-order Runge-Kutta method for the derivatives. For a given $\ell$ value, a vortex is found by tuning $a_{\ell-1}$ towards a critical value $a_{\ell-1}^{\textrm{vortex}}$. For instance, for the three lowest rotational values (nonzero rotation in both spinor components), we found 
\begin{eqnarray}
a_1^{\textrm{vortex}} &=& 0.571718...  \; \;, \hspace{3pc} \ell =2 \, ,  \\
a_2^{\textrm{vortex}} &=& 0.145291...  \; \; , \hspace{3pc} \ell =3 \, ,\\
a_3^{\textrm{vortex}} &=& 0.0240267...  \;\;  , \hspace{2.6pc} \ell =4   \, .
\end{eqnarray}
Figure~\ref{NumVortexConv} displays the shooting process for radial profiles in the case $\ell =2$. We have used the same shooting method to obtain the ring-vortex solutions ($\mu=0$ with asymptotically vanishing tails), in addition to the exact algebraic closed forms in Table~\ref{table2}.

To compute vortex lifetimes requires a framework analogous to the Bogoliubov-de Gennes system but tailored to the particular structure of the NLDE. The RLSE provide this framework forming a relativistic generalization of the Bogoliubov-de Gennes equations analogous to the relationship between the NLDE and nonlinear Schr\"odinger equation. Thus, in the RLSE the quasi-particle amplitudes $u$ and $v$ are each vector in form, to match the four-spinor (two-spinor at one Dirac point) they perturb from. The RLSE can be expressed in $2 \times 2$ matrix-vector form:
\begin{eqnarray}
\hspace{-2pc}\tilde{\mathscr{D}}  {\bf u}_{\bf k}  - U \tilde{{\Psi}}   {\bf v}_{\bf k} &=&  \tilde{ E}_{\bf k} {\bf u}_{\bf k} , \label{eqn:RLSE5} \\
\hspace{-2pc}\tilde{\mathscr{D}}^*  {\bf v}_{\bf k}  - U \tilde{{ \Psi }}  {\bf u}_{\bf k} &=& - \tilde{E}_{\bf k} {\bf v}_{\bf k}  \; ,  \label{eqn:RLSE6}\end{eqnarray} 
where $\tilde{\mathscr{D}}$ and $\tilde{{\Psi}}$ are $2 \times 2$ matrices which contain the first-order derivatives $(\partial_x + i \partial_y)$ and the background BEC components $\psi_A$, $\psi_B$, and $\tilde{ E}_{\bf k}$ is the $2 \times 2$ eigenvalue matrix. Note that $U$ is the particle interaction. When broken down, Eqs.~(\ref{eqn:RLSE5})-(\ref{eqn:RLSE6}) form a $4 \times 4$ eigenvalue problem in the quasi-particle amplitudes $u_{k, A(B)}$ and $v_{k, A(B)}$ (with momentum ${\bf k}$) associated with particle and hole excitations of the A(B)-sublattices at a Dirac point. Vortices possess cylindrical symmetry so we express Eqs.~(\ref{eqn:RLSE5})-(\ref{eqn:RLSE6}) in plane-polar coordinates, factor the quasi-particle amplitudes into radial and angular parts, then substitute in the particular solution for $\psi_{A(B)}$. We then obtain a set of first-order coupled ODE's in the radial coordinate to be solved consistently for the functions $u_{A(B)}(r)$, $v_{A(B)}(r)$ and the associated eigenvalues. We discretize the derivatives and functions using a forward-backward average finite-difference scheme, then solve the resulting discrete matrix eigenvalue problem using a standard numerical diagonalization method.

To compute vortex lifetimes, we solve the RLSE to obtain the quasi-particle spatial functions and eigenvalues. In general, for vortex solutions of the NLDE certain eigenvalues and eigenmodes key to understanding the physical motion correspond to Nambu-Goldstone modes, i.e., anomalous with a small imaginary component~\cite{feder2000}. When thermal losses are small, it is the imaginary part of the linear eigenvalues which depletes the BEC. We define the vortex lifetime by computing the time for depletion to reach a significant fraction of the total fixed number of atoms in the system, and consider only depletion coming from the mode with the largest imaginary term in its eigenvalue. The lifetime is then given by $\tau = \left[  \hbar /\mathrm{Im}(E) \right] \mathrm{ln} \left(R_\perp/I \right)$, expressed in terms of the largest linear eigenvalue $E$ and the planar radius of the BEC $R_\perp$, in units of the lattice constant $a$ (see Table~\ref{table1}). Note also that the spatial integral $I$ here is specific to each vortex type and involves overlaps of the quasi-particle and condensate spatial functions. For the experimental parameters of Table~\ref{table1}, we find the longest lived solutions to be the vortex/soliton and Anderson-Toulouse vortex with $\tau = 11.51 \, \mathrm{s}$, compared to the typical lifetime of a $^{87}\textrm{Rb}$ condensate in an optical lattice of less than a second~\cite{trotzky2010}.

For most of vortex types (i)-(vii), we find lifetimes $\tau$ to be long compared to the lifetime of the BEC itself. In particular, we obtain the following values for $\tau$: $9.13 \, \mathrm{s}$, $10.43 \, \mathrm{s}$, $11.51 \, \mathrm{s}$, $1.57\!  \times \!10^{-7} \, \mathrm{s}$, $1.57\! \times \!10^{-7} \, \mathrm{s}$, $1.25 \, \mathrm{s}$, $1.29 \! \times \! 10^{-5} \, \mathrm{s}$; for the vortex/soliton, ring-vortex/soliton, Anderson-Toulouse, Mermin-Ho, half-quantum, $\ell =2$ ring-vortex, and $\ell =2$ topological vortex, respectively.

\begin{figure}[h]
\begin{center}
\subfigure{\hspace{-.2in} \includegraphics[width=.177\textwidth,height = .115\textheight]{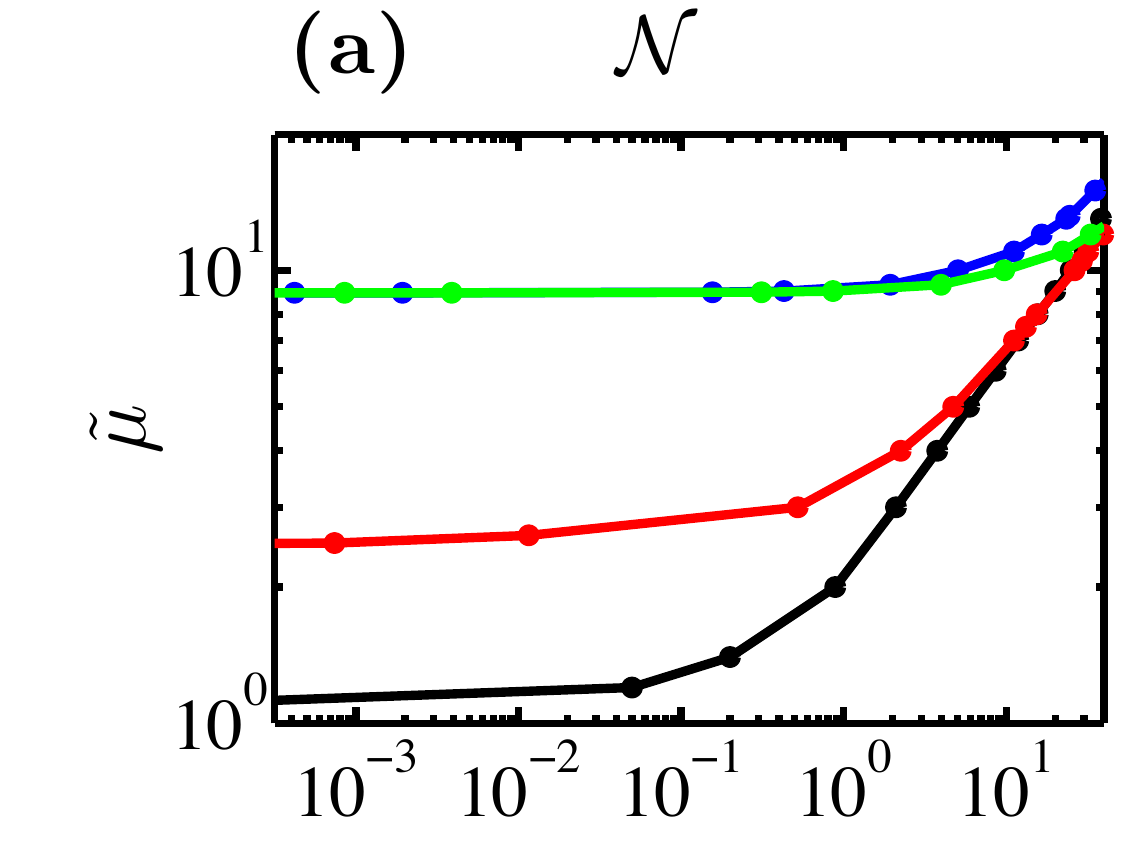} \hspace{-.08in}
 \includegraphics[width=.177\textwidth,height = .115\textheight]{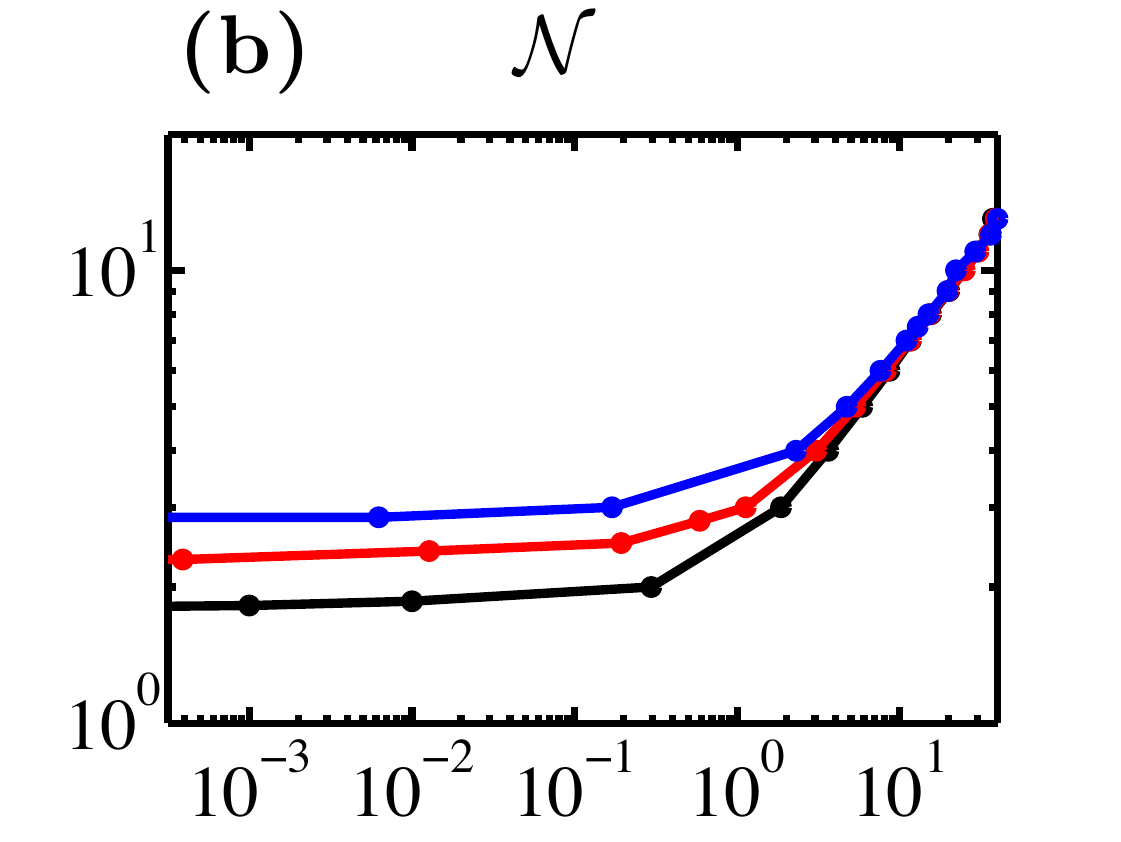} 
\hspace{-.15in} \includegraphics[width=.18 \textwidth,height = .118 \textheight]{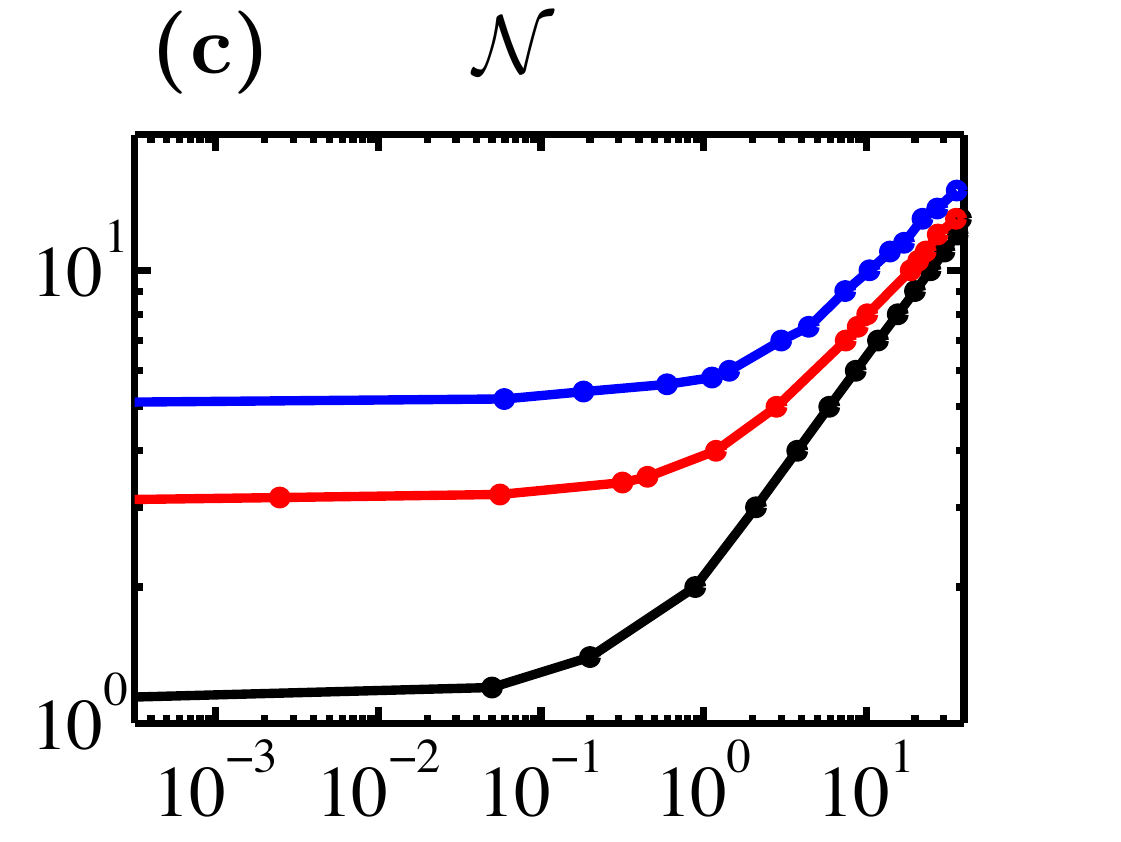} } \\
   \end{center}
\caption[]{(color online) \emph{Spectra for relativistic vortices confined in a harmonic potential}.(a) Vortex/soliton (black curve), Anderson-Toulouse skyrmion (red), Mermin-Ho skyrmion (blue), and half-quantum vortex (green). (b) Topological vortices for $\ell=2, 3, 4$ (black, red, blue). (c) Radial ground state and first two excited states of the vortex without skyrmion symmetry (black, red, blue). In each figure, the renormalized chemical potential is plotted as a function of the normalization. There are two regimes characterized by power laws: $\tilde{\mu} \propto \mathcal{N}^\alpha$. The weakly interacting free-particle regime occurs for small $\mathcal{N}$, whereas the strongly interacting vortex regime is in the region of large $\mathcal{N}$. Note that the vertical and horizontal axes labels are dimensionless. }
\label{cnfnd3}
\end{figure}

In order to have a clear comparative prediction for energies involved in creating our vortices, we solve the NLDE using a numerical shooting method in the presence of a weak harmonic trap of frequency $\omega_\perp = 2\pi \times 0.0387 \, \mathrm{Hz}$ along the direction of the lattice. This is the frequency associated with a planar BEC radius equal to $100$ times the lattice constant. In this case vortices come in radially quantized states. For simplicity, we focus mainly on the lowest radial excitation. Using a generalization of the method in~\cite{CarCla06}, we have obtained the dimensionless (renormalized) chemical potential $\tilde{\mu} \equiv  \mu/\hbar \omega_\perp$ as a function of the normalization $\mathcal{N} =  \sqrt{3}\, \hbar \omega_\perp  N   U/ 3  t_h^2$ for each vortex type, as shown in Fig.~\ref{cnfnd3}. Here, $N$ is the number of atoms in the system with the other quantities defined in Table~\ref{table1}. Note that ring-vortices are minimally affected by the presence of a weak trap, since they are highly localized objects and lie very near the center of the trap.

\section{Experimental realization of vortices}
\label{sec:CreateVortex}

In this section, we discuss how relativistic vortex solutions of the NLDE can be excited by modifying the technique for coherent sublattice transfer described in Sec.~\ref{sec:SublatticeTransfer}. Starting from a condensate at the Dirac point with $m_F = 0$, and non-zero amplitude in sublattice A only, a vortex excitation can be created by replacing the second microwave transition shown in Fig.~\ref{fig:SublatticeTransfer} with a two-photon Raman transition with one of the photons carrying a single unit of orbital angular momentum. The two-photon Raman transition drives Rabi oscillations between two hyperfine states in the electronic ground state of an atom by coupling through intermediate states which are optically excited electronic states. The transition matrix element between hyperfine states is proportional to the product of the two field amplitudes which drive the two-photon transition.  To excite a vortex, the two optical fields are provided by co-propagating Gaussian and Laguerre-Gaussian laser beams which have a frequency difference corresponding to the energy splitting between the initial and final states but are both far-detuned from the intermediate states to reduce spontaneous emission. The Laguerre-Gaussian beam carries a single unit of orbital angular momentum which is transferred to the atoms in the stimulated Raman transition~\cite{Phillips06}. The electric field amplitude of a Laguerre-Gaussian laser beam with radial mode index $p = 0$ and charge index $\ell = 1$ is proportional to
\begin{eqnarray}
E_\mathrm{LG}^{p=0,\ell = 1} (r,\theta) \propto r \, \exp\left( - \frac{r^2}{w_0^2} \right) \, \exp(i \theta) \, ,
\end{eqnarray}
where $r$ and $\theta$ are respectively the radial and azimuthal coordinates relative to the optical axis and $w_0$ is the beam waist.  The field of the Gaussian laser beam $E_\mathrm{G} (r,\theta) \propto \exp\left( - r^2/w_0^2 \right)$. Thus, the effective Rabi frequency for the two photon transition $\Omega_{2 \gamma} \propto \left\langle f \right| E_\mathrm{G}(\mathbf{r}) \, E_{\mathrm{LG}}^{p=0,\ell =1}(\mathbf{r}) \left| I_2 \right\rangle$ where $\left| I_2 \right\rangle$ and $\left| f \right\rangle$ are respectively the intermediate and final state spatial wavefunctions of the condensate depicted in Fig.~\ref{fig:SublatticeTransfer}.  Due to the azimuthal phase winding $\exp(i \theta)$ of the LG field $E_{\mathrm{LG}}^{p=0, \ell=1}$, the Raman fields provide the appropriate spatial dependence to drive a transition to a final state $\left| f \right\rangle$ which has a single unit of angular momentum starting from the intermediate state $\left| I_2 \right\rangle$ with no orbital angular momentum.

Starting from a condensate at the Dirac point ${\bf{K}}$ with amplitude only in the A sublattice sites, i.e., the Bloch state $\psi_{A,\mathbf{K}}$, the procedure described above would couple to a vortex/soliton solution of the NLDE which has a vortex in the B sublattice and a soliton, with no angular momentum, in the A sublattice. This solution of the NLDE in the continuum limit can be written as a Weyl spinor of the form $\Psi_f = (\psi_A, \psi_B) = \left[ i f_A(r), f_B(r) \exp(i \theta) \right]$ (see Ref.~\cite{haddad2011}). The initial wavefunction of the condensate at the Dirac point $\psi_{A,\mathbf{K}}$ is described by the Weyl spinor $\Psi_i = (\psi_A, \psi_B) =  (1,0)$. In the transition sequence depicted in Fig.~\ref{fig:SublatticeTransfer}, the condensate initially in the state $\left| i \right\rangle = \psi_{A,\mathbf{K}}^{m_F = 0}$ is transferred via a mw field to an intermediate state with $m_F = 1$ at the Dirac point of the A sublattice (i.e. $\left| I_1 \right\rangle = \psi_{A,\mathbf{K}}^{m_F = 1}$), subsequently transferred to the B sublattice (i.e. $\left| I_2 \right\rangle = \psi_{B,\mathbf{K}}^{m_F = 1}$) by modulation of the lattice potential through application of $H_m \, \cos\omega_m t$, and ultimately transferred by the two-photon Raman transition to the final state $\left| f \right\rangle$ which is the vortex/soliton state in the internal state with $m_F = 0$. If we assume that $w_0, \xi \gg a$ and take the tight binding and continuum limits, the effective Rabi Raman frequency
\begin{eqnarray}
\Omega_{2 \gamma} & \propto & \left\langle I_2 \right| E_{\mathrm{G}}(\mathbf{r}) \, E_{\mathrm{LG}}^{p=0, \ell=1}(\mathbf{r}) \left| f \right\rangle \\ & \propto & E_{\mathrm{G},0} \, E_{\mathrm{LG},0} \, \int f_B(r) \, r^2 \, e^{-2 r^2/w_0^2} \, dr. \nonumber
\end{eqnarray}
The radial dependence of the vortex in the B sublattice $f_B(r)$ was calculated in our previous work~\cite{Haddad2012}. The radial integral is positive definite and for $w_0 \sim \xi$ will give a non-zero Rabi frequency with an absolute value determined by the amplitudes of the fields driving the two-photon Raman transition and the dipole transition matrix elements for the 5S-5P electronic transitions in $^{87}$Rb.

In order to apply our discussion to specific vortex types, we first consider the excitation of a relativistic vortex starting with all the atoms in the A sublattice at the Dirac point. We then apply the co-propagating Gaussian and Laguerre-Gaussian laser beams, as explained. The spatial variation of the beam results in mainly the B sublattice being populated (the vortex) throughout most of the 2D lattice, except within a small disk which becomes the core of the vortex. On the other hand, the A sublattice is left depleted everywhere except near the core of the vortex (the soliton). This describes excitation of the vortex/soliton or Anderson-Toulouse skyrmion~\cite{haddad2011}. The Mermin-Ho vortex can be obtained by the same process, but by only partially transferring atoms to the B sublattice. The sublattice amplitudes far from the vortex core are tuned to satisfy $|\psi_B|^2 = |\psi_A|^2 < 1$, where $|\psi_{A(B)}|^2$ is the density of the BEC in the first (second) four-spinor component in the NLDE, and $\mathbf{v}_{A(B)} = (\hbar/M) \nabla \phi_{A(B)}$ is the associated relativistic fluid velocity, with $\phi_{A(B)}= \textrm{Arg}(\psi_{A(B)})$ the phase. The half-quantum vortex or semion can be excited by using a fractional optical vortex beam in order to provide the required angular phase jump~\cite{Leach2004,Basistiy2004}. General topological vortices have phase winding $\ell >1$, non-zero chemical potential $\mu$, and satisfy $|\psi_A|,  |\psi_B| \ne 0$ far from the center of the trap. General topological vortex excitations may be induced by subsequent applications of a two-photon transition with co-propagating Laguerre-Gaussian/Gaussian beams which transfer the condensate between $m=0$ states (i.e. from $F=1, \, m=0$ to $F=2, \, m=0$ or vice versa). Each two-photon transition changes the orbital angular momentum of both the A and B sublattices by the orbital angular momentum carried by the Laguerre-Gaussian beams, while maintaining the desired winding differential between the A and B sublattices. Finally, ring-vortices, characterized by $\mu =0$ and $|\psi_A|,  |\psi_B| = 0$ far from the center of the trap, can be obtained from the other vortices by inducing depletion of the BEC from the outer edge of the trap towards the core. More details regarding solutions of the NLDE may be found in Ref.~\cite{haddad2011}.

\section{Conclusion}
\label{sec:Conclusion}

In conclusion, we have described in detail a method for constructing a stable BEC at the Dirac points of a honeycomb optical lattice. Our system allows for relativistic vortex excitations in a macroscopic Dirac spinor wavefunction, providing a means of studying high energy field theoretic vortices in a condensed matter setting. We have completely specified the required physical parameters, lifetimes, and spectra for harmonically bound vortices as a prescription guide for the experimentalist. Variations on the NLDE have tremendous potential for a host of relativistic simulations in BECs. Interesting examples include Soler models~\cite{Soler1974} and the extended Gross-Neveu model~\cite{Lee1975}.  Our work puts such efforts on a solid experimental footing.

\section*{ACKNOWLEDGMENTS}
This material is based in part upon work supported by the National Science Foundation under grant numbers PHY-1207881, and the Air Force Office of Scientific Research grant number FA9550-08-1-0069. L.D.C. thanks the Alexander von Humboldt foundation and the Heidelberg Center for Quantum Dynamics for additional support.

\bibliographystyle{prsty}
\bibliography{NLDE_Vortex_Refs}

\section*{APPENDIX A: Convergence of solutions used to compute spectra for radially confined vortices}

\begin{figure}[t!]
\centering
\vspace{0pc}
\hspace{-.075in}\subfigure{
\label{fig:ex3-b}
\includegraphics[width=.475\textwidth]{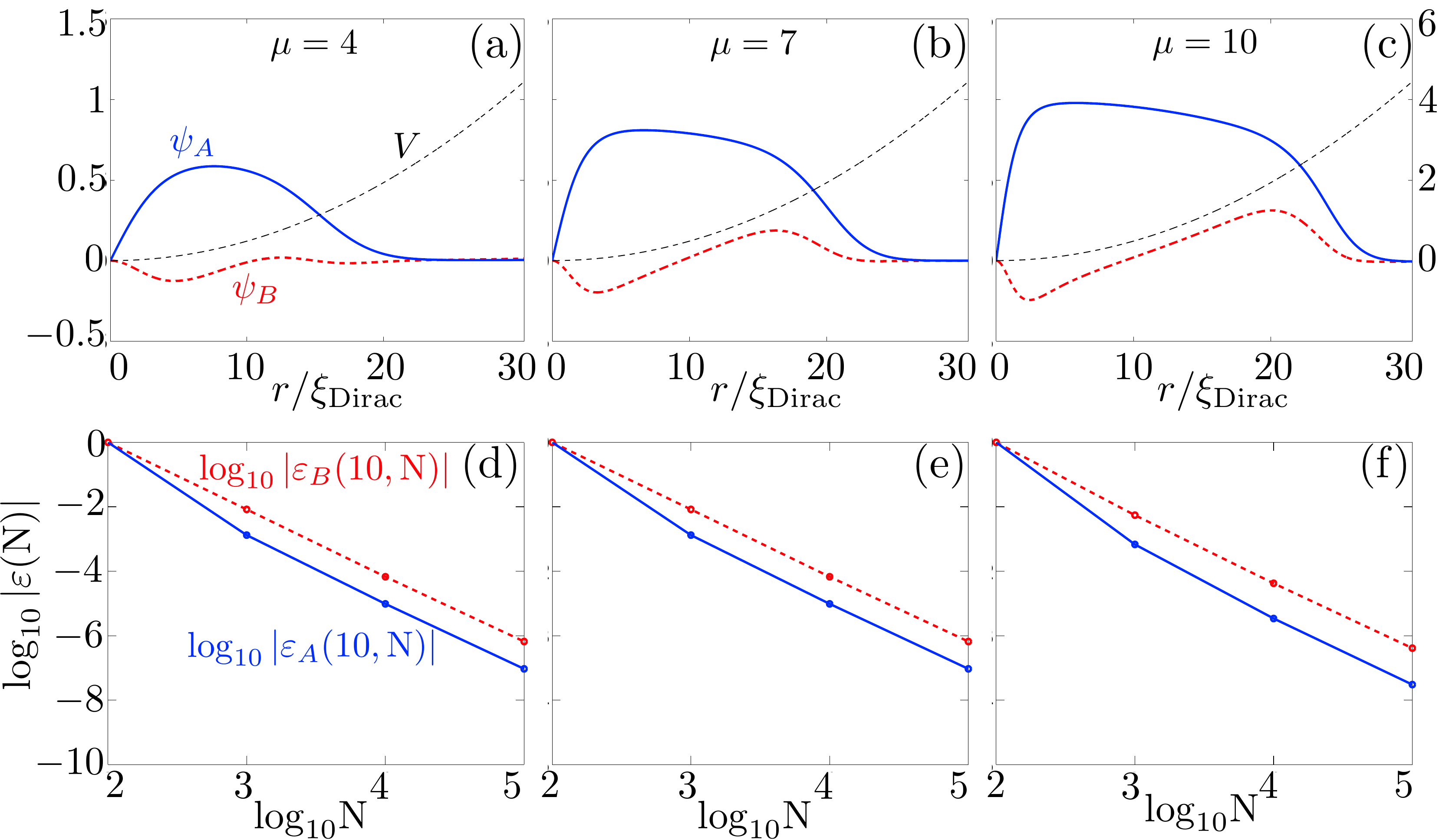}}   \\
\caption[]{(color online) \emph{Convergence of $\ell =2$ topological vortex radial profiles}. (a)-(c) The explicit radial profiles for $U= 1$, $\mu =4, \, 7, \, 10$ and grid size $\mathrm{N} = 10^6$. The black dashed curve is the harmonic potential. The scale for the potential is shown on the right hand vertical axis of panel (c) in units of $\mathrm{nK}$. (d)-(f) Log-log error profiles computed using Eq.~(\ref{error}). Note that the curves are a guide to the eye with data points representing actual data. }
\label{VortWithConv}
\end{figure}

To show convergence of the radial ground state of the $\ell =2$ vortex in a harmonic trap, we focus on three of the solutions which make up the black curve in Fig.~\ref{cnfnd3}(b). The radial profiles of these solutions, $\psi_A$ and $\psi_B$, are shown in Figs.~\ref{VortWithConv}(a)-(c) and correspond to the chemical potentials $\mu = 4$, $\mu = 7$, and $\mu = 10$ interpolating between the free-particle and strongly nonlinear limits, respectively. These solutions were obtained by finite differencing using a shooting method to tune the precision of the initial value of $\psi_A$ such that $\psi_A \ll 1$ to pick out the ground state. For convergence at a single radial point, we compute the value of the solution at the dimensionless radius $\chi_i \equiv r_i/\xi_{\mathrm{Dirac}} = 10$ for several values of the grid size $\mathrm{N} = 10^2, \, 10^3, \, 10^4, \, 10^5, \,10^6$. We use the error formula which depends on the dimensionless radius and number of grid points 
\begin{eqnarray}
\varepsilon_{A(B)}(\chi_i, \mathrm{N})  \equiv \left[ \frac{  \psi(\chi_i)_{A(B)}^{\mathrm{N}+1} - \psi(\chi_i)_{A(B)}^\mathrm{N} }{ \psi(\chi_i)_{A(B)}^{\mathrm{N}+1} + \psi(\chi_i)_{A(B)}^\mathrm{N}  }\right] \, , \label{error}
\end{eqnarray}
where in the symbol $\psi(\chi_i)_{A(B)}^{\mathrm{N}}$ the subscript $A(B)$ denotes the sublattice excitation, $\chi_i$ denotes the $i^{\mathrm{th}}$ element in the discretized dimensionless radial coordinate, and the superscript $\mathrm{N}$ denotes the number of grid points used in the calculation. In Figs.~\ref{VortWithConv}(d)-(f), we have plotted $\mathrm{log}_{10}\left| \varepsilon_{A(B)}(\mathrm{10,N}) \right|$ versus $\mathrm{log}_{10}\mathrm{N}$, for the solutions shown in Figs.~\ref{VortWithConv}(a)-(c).

\end{document}